\documentclass[10pt,a4paper]{article}
\usepackage{graphicx,amsmath,color,psfrag,epsfig,rotate,rotating,epsf}

\newcommand{\pl}{P_{\tr{left}}}
\newcommand{\pmm}{P_{\tr{middle}}}
\newcommand{\tr}{\textrm}
\newcommand{\h}{\frac{1}{2}}
\newcommand{\g}{\Gamma}

\newcommand{\dd}{\textrm{d}}
\newcommand{\la}{\langle}
\newcommand{\ra}{\rangle}
\newcommand{\re}{\textrm{Re}}
\newcommand{\im}{\textrm{Im}}

\begin{document}
\parindent 0mm
\parskip 6pt
\title{SLE in the three-state Potts model - a numerical study}
\author{Adam Gamsa and John Cardy\\
Rudolf Peierls Centre for Theoretical Physics\\
         1 Keble Road, Oxford OX1 3NP,
         U.K.}
\date{April 2007}
\maketitle
\begin{abstract}
The scaling limit of the spin cluster boundaries of the Ising model with domain wall
boundary conditions is
SLE with $\kappa=3$. We hypothesise that the three-state Potts model with appropriate
boundary conditions has spin cluster boundaries which are also SLE in the scaling limit,
but with $\kappa=10/3$.
To test this, we generate samples using the Wolff algorithm and test them against
predictions of SLE: we examine the statistics of the Loewner driving function, estimate the fractal
dimension and test against Schramm's formula. The results are in support of our hypothesis.
\end{abstract}
%

\section{Introduction}\label{SecIntroduction}
Much of the recent interest in Schramm Loewner evolution (SLE) stems from the
description it offers of cluster boundaries in critical
statistical mechanics models. These models are conjecturally described by
conformal field theories, which determines expectation values
of products of local operators. However, non-local objects
such as cluster boundaries do not have a natural expression in the
field theory language. These curves are of great
interest as, for example, they describe percolation cluster
boundaries, level lines of height models and spin cluster
boundaries. Recently, the scaling limit of
the spin cluster boundary in the Ising model with appropriate
boundary conditions has been proven to be SLE with
$\kappa=3$~\cite{SmirnovIsing}.

The Ising model corresponds to the $q$-state Potts model with
$q=2$. We may ask, therefore, whether spin boundaries in
the critical Potts model with other values of $q$ have
SLE as their scaling limit.
The Fortuin-Kastelyn (FK) cluster representation~\cite{FKcluster} of the Potts model
contains closed loops which may be taken to be the contours of a
suitably defined height model.
The scaling limit of this model is supposed to be a field theory (CFT) called the
Coulomb gas which possesses a single parameter $g$, where $\sqrt{q}=-2\cos(\pi g)$.
The connection with SLE is that the winding angle of these loops in
the Coulomb gas are equivalent to that seen in radial SLE for
$\kappa=4/g$. Therefore, we have reason to believe that the FK
cluster boundaries should correspond to SLE in the scaling limit,
with
\begin{equation}\label{EqnKappa}
 \kappa=\frac{4}{1-\arccos(\sqrt{q}/2)/\pi}\,.
\end{equation}

The spin cluster boundaries are related to these FK cluster
boundaries in a non-trivial way, however, and it is therefore
interesting to investigate the properties of these spin boundaries
in the scaling limit. For the Ising model, these spin cluster
boundaries are also described by SLE at the dual value $\kappa'=16/\kappa$~\cite{SmirnovIsing}.

Besides $q=1$ (corresponding to percolation), there are two other
integer values of $q$ for
which the Potts model displays a continuous phase transition: $q=3$
and $q=4$ (for $q>4$ the transition is first order). In this paper,
we investigate the case $q=3$ and test for agreement of
numerically simulated lattice spin cluster boundaries with known predictions of
SLE. The dual value of $\kappa$ for $q=3$ is, from
equation~\ref{EqnKappa} above, $\kappa'=16/\kappa=10/3$. We may
hypothesise, therefore, that the spin cluster boundaries, properly defined, of the
$q=3$ Potts model are SLE with $\kappa=10/3$. This paper is a
numerical investigation of that prediction.

The set-up of chordal SLE requires that the curves connect fixed points on the boundary
of a simply connected domain. This is ensured by careful choice of the boundary conditions.
For example, consider the following boundary conditions for the Ising model:
divide the boundary into two connected subsets (`left' and `right').
Let the left boundary spins be spin up and the right boundary spins
be spin down.
This ensures that a spin cluster boundary runs from one
fixed point on the boundary to another, for every configuration of
the bulk spins. The Gibbs distribution induces a measure on these
curves. The case of the $q=3$ Potts model is similar,
although there are now two types of interesting boundary
conditions. If the boundary conditions are chosen as for the Ising
model, with those on the left part being of spin type~$1$ (say) and those on the right of spin
type $2$, the right boundary curve of the cluster connected to the
boundary spins of type~$1$ does not everywhere coincide with the
right boundary curve of those of type~$2$. In other words, the spin cluster
boundary may split around regions of spin type~$3$, which themselves may
contain islands of the other spin types, ad infinitum. One possibility for the scaling limit is
that the two cluster boundaries are distinct, with probability one.
This might then be expected to correspond to two-curve
SLE. We therefore simulate numerically a set of
samples with these boundary conditions and compare to predictions
from two-curve SLE: the fractal dimensions and the 
formulae describing the expected spatial distribution of the
curves in the bulk~\cite{CardyGamsa}. The other choice for boundary conditions is to
fix the left set of the boundary spins to
be of type~$1$ and to allow those on the right to fluctuate
between the other two spin types. In this case, we may consider the right
boundary of the cluster connected to the boundary
spins of type~$1$, which forms a curve connecting the two fixed
points on the boundary. The measure on this set of curves is, as before,
induced by the Gibbs distribution. We also simulate
Potts models with these boundary conditions to test against two
predictions of single-curve SLE: the fractal dimension and Schramm's
formula. For the single curve case, we also examine the statistics of the driving function
by undoing the sequence of conformal
transformations which uniquely determine the form of the curve
through the Loewner equation. This driving function should be
Brownian motion with diffusivity $\kappa$ if the curves correspond to
SLE.

The layout of the paper is as follows: In section~\ref{SecPotts},
we summarise the definition and important features of the Potts
model, explaining in detail the boundary conditions required to
generate cluster boundaries. In section~\ref{SecSLE}, we
recall the definition of SLE and describe the predictions of SLE
which we will test our samples against. In
section~\ref{SecSimulations} we describe the details of the
numerical simulations we performed, explaining how the fractal
dimension and other data are extracted from the
generated samples. In section~\ref{SecResults}, we present the
results of our numerical simulations and conclude in
section~\ref{SecSummary}. The appendix contains details of the
Wolff algorithm used for all simulations.

\section{The Potts model}\label{SecPotts}
The $q$-state Potts model is a lattice spin model. The spins may
take values from the set $\{1,2...,q\}$ and have nearest neighbour
interactions, such that the partition function takes the form
\begin{equation}
Z=\tr{Tr}\, e^{J\,\sum_{r,r'}{\delta_{s(r),s(r')}}}\,.
\end{equation}
$J$ is the reduced coupling constant and $\delta_{i,j}$ is the
Kronecker delta. The case $q=2$ is equivalent to the Ising model,
up to a constant. For $q>2$, therefore, the Potts model describes
a generalisation of the Ising model to systems with more than two
spin types. For a given $q\leq 4$ and choice of
lattice, there exists a value of the coupling constant, $J_{c}$,
at which the model has a continuous transition; the two point correlation
function of the spins decays as a power law and the scaling limit
of the model is supposed to be described by a
conformal field theory. For later reference, these values for the
two lattice types considered in this work are~\cite{Wu}
\begin{itemize}
\item Square lattice $e^{J_{c}}=\sqrt{q}+1$.
\item Triangular lattice $e^{J_{c}}=2\cos(\frac{2}{3}\cos^{-1}(\sqrt{q}/2))$.
\end{itemize}

\begin{figure}[thbp]
  \begin{center}
  \scalebox{0.5}{\centerline{
   \epsfig{figure=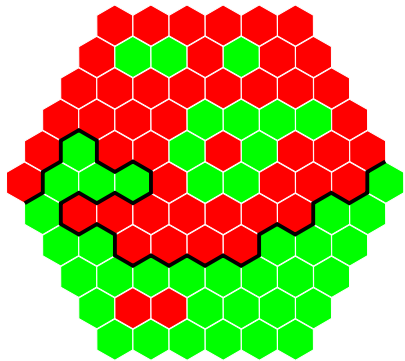}}}
    \caption{The Ising model with boundary conditions as in the text.
The spins are represented by hexagons, so that there are six
nearest neighbours for each spin in the bulk. Half of the boundary
spins are of type $1$ (coloured red, or dark grey on black and white printouts)
and the others of type $2$ (green or light grey).
There exists a boundary of the cluster connected to the bottom
edge (the thick black line) which coincides with the boundary of
the cluster connected to the top edge. This curve connects the
same two points on the boundary for every configuration of the
bulk spins.}
    \label{FigIsBound}
  \end{center}
\end{figure}

Let us consider spin cluster boundaries in the Ising model arising
from a specific choice of boundary conditions. Namely, let those
spins on a connected piece of the boundary be of spin type~$1$ and
elsewhere on the boundary, the spins be of type~$2$. This is
demonstrated in figure~\ref{FigIsBound}, where spin type~$1$ is
represented by the colour red and type~$2$ spins are coloured
green. Then, for every configuration of the spins in the bulk,
there exists a unique curve starting and ending at two fixed
points on the boundary. Immediately to one side of the curve, all
the spins are green. To the other side, the spins are red. The
measure on these curves is derived from the Gibbs distribution.


\begin{figure}[thbp]
  \begin{center}
    \begin{minipage}[c]{0.5\linewidth}
      \scalebox{0.8}{\epsfig{file=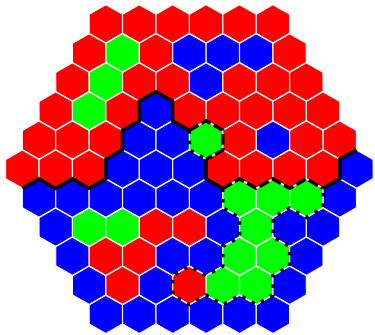, width=\linewidth}}
    \end{minipage}\hfill
    \begin{minipage}[c]{0.46\linewidth}
      \caption{The $q=3$ Potts model with `fixed' boundary conditions. The clusters
connected to the bottom and top edges are no longer everywhere
overlapping. The dashed and dotted lines are `split' cluster
boundaries. The thick black line is the `composite' cluster
boundary.\label{Figq3bBound}}
    \end{minipage}
  \end{center}
\end{figure}

Let us consider how to define the boundary conditions for the
$q=3$ Potts model in such a way as to guarantee the existence of
cluster boundary curves propagating through the sample. The
natural extension of the boundary conditions for the Ising model
is to fix a connected set of the boundary spins to be of (say) spin type~$1$
(red or dark grey on black and white printouts), with the remaining spins
of type $3$ (blue or black), as in
figure~\ref{Figq3bBound}. The particular choice of spins is, of
course, arbitrary due to the $S_{3}$ symmetry of the model. We
shall refer to these boundary conditions as `fixed' boundary
conditions. However, the situation is different from the Ising
model, since in this case the boundary curve of the cluster
containing the red boundary spins does not always coincide with
the boundary of the cluster containing the blue boundary spins. In
the figure, the dotted line is the boundary of the cluster
containing the blue boundary spins, the dashed line is the
boundary of the cluster containing the red boundary spins and
where the two curves are both present along an edge, a thick black
line has been drawn. This black line is reminiscent of the single
cluster boundary in the Ising model, with spins immediately to one
side of it red and those on the other side being blue, but the
dotted and dashed lines are of a new type. The dotted line has
blue spins to one side and either green (light grey on black and white printouts)
or red to the other. The
dashed line has red spins to one side and either green or blue to
the other side. In this paper, we shall refer to the thick black
lines as the `composite' cluster boundaries and the dashed and
dotted lines as the `split' cluster boundaries. Note that the
`composite' cluster boundary splits into two `split' cluster
boundaries whenever it encounters a vertex with all three spin
types, and that a pair of `split' cluster boundaries recombine
only at such three spin vertices.

\begin{figure}[thbp]
  \begin{center}
    \begin{minipage}[c]{0.5\linewidth}
      \scalebox{0.8}{\epsfig{file=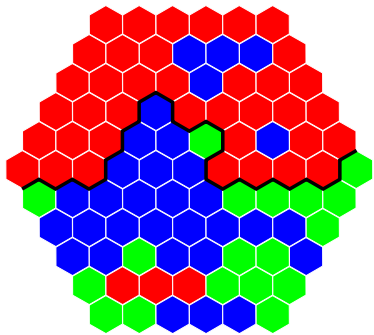, width=\linewidth}}
    \end{minipage}\hfill
    \begin{minipage}[c]{0.46\linewidth}
      \caption{The $q=3$ Potts model with `fluctuating' boundary conditions.
There exists a single uniquely defined boundary curve between the
red (dark grey) cluster and the green (light grey) and blue (black) cluster, as
defined in the text.\label{Figq3Bound}}
    \end{minipage}
  \end{center}
\end{figure}

There exists another choice of boundary conditions for the $q=3$
Potts model which we will refer to as `fluctuating' boundary
conditions. A connected set of the boundary spins are set to be of spin type
$1$ (red). The remaining boundary spins are allowed to fluctuate
with the spins in the bulk, but are not permitted to be of
type~$1$. They are therefore of type~$2$ (green) or $3$ (blue),
see figure~\ref{Figq3Bound}. The cluster boundary of the red spin
cluster connected to the top edge now coincides with the boundary
of a cluster comprised of green and blue spins connected to the
bottom edge, which is defined as follows: replace all green and
blue spins with a new spin type $x$ and consider the boundary of
the cluster of $x$ spins connected to the bottom edge. That these
two cluster boundaries always coincide may be readily seen by
analogy with configurations of the Ising model with spin types red
and $x$. The reader will note the lack of reflection symmetry in the boundary conditions.
This may be expected to lead to non-reflection symmetric
spin cluster boundary curves. However, we shall present evidence that these curves
are actually reflection symmetric in the scaling limit.

\section{Schramm Loewner evolution}\label{SecSLE}
We expect the cluster boundary curves of critical statistical
models to be described by measures which are invariant under
conformal transformations, due to the conformal invariance of the
action associated to these models. This argument can be made more
precise; the changes in boundary conditions described in the previous section
have been shown~\cite{CardyBoundary,SaleurBauer} to be produced by
the insertion of boundary $\phi_{1,2}$ operators in the
appropriate conformal field theory. Expectation values of
observables therefore correspond to correlation functions with a
pair of these boundary condition changing operators inserted. It
may be shown, for a specific choice of observables, that the
partial differential equations for these correlation functions
arising from the null state associated with the $\phi_{1,2}$
operators are the same as those coming from the mathematical
theory of random curves with conformally invariant measure, known
as Schramm Loewner evolution (SLE).

Let us summarise the main features of the theoretical physicist's
version\cite{CardyRev,BauerBernard} of SLE. A curve in the upper
half plane from the origin to infinity is considered as being
grown dynamically as a function of time $t$, which therefore
parameterizes distance along the curve. Let $K_t$ be the set
consisting of the curve as grown up to time $t$ (as well as all
points enclosed by the curve and between the curve and the real
axis) so that the complement of this set in the upper half plane
is simply connected. Let $g_t(z)$ be the (unique) conformal
mapping of this complement to the whole upper half plane,
normalised such that $g_t(z)=z+O(1/z)$ as $z\to\infty$. The
coefficient of the $O(1/z)$ term is increasing with $t$, so `time'
can be reparametrised such that this coefficient is exactly $2t$.
Then, $t$ is known as the capacity. The image of the growing tip
of the curve under $g_t$ is a point $a_t$ on the real axis.
Loewner showed that the time-evolution of $g_t$ satisfies a
differential equation, which is named after him:
\begin{equation}\label{EqLoewner}
\frac{dg_t(z)}{dt}=\frac2{g_t(z)-a_t}\,.
\end{equation}
Any suitably continuous function $a_t$ generates a curve.
Schramm\cite{SchrammIsrael} showed that for this process to
generate a conformally invariant measure on curves, the driving
function must be $a_t=\sqrt\kappa B_t +a_0$, with $B_t$ being a
standard Brownian motion.

There remains a single parameter in the theory, $\kappa$, which is
the diffusivity of the Brownian motion. A small value of $\kappa$
results in a measure which favours curves with few turns, whereas
a large value of $\kappa$ leads to a measure favouring curves
which wind more frequently, touching themselves and the real axis.
The value of $\kappa$ is conjecturally related to the central
charge of the conformal field theory it describes via the formula:
\begin{equation}\label{Eqckappa}
c=\frac{(3\kappa-8)(6-\kappa)}{2\kappa}\,.
\end{equation}
From this formula, it is clear that the central charge associated to an SLE
with diffusivity $\kappa$ is equal to that of the dual SLE with $\kappa'=16/\kappa$.

We will now describe some properties of SLE curves.

\subsection{The fractal dimension}\label{SubSecFract}
If $N(r)$ is the number of discs of radius $r$ required to cover
an object, this number scales as $N(r)\sim r^{-d_{f}}$ for small
$r$, where the exponent $d_{f}$ is known as the fractal dimension.
SLE curves are fractal objects, with
\begin{equation}\label{EqFract}
  d_{f}=1+\frac{\kappa}{8}\,,\quad\textrm{for}\,\kappa\leq 8\,.
\end{equation}

The total length of the curve, $S$, measured in units of the lattice spacing, $a$,
obeys a scaling relation as a function of system size, $L$, as $L/a\rightarrow\infty$
\begin{equation}\label{EqFractLength}
S\sim a \left(\frac{L}{a}\right)^{d_{f}}\,.
\end{equation}

\subsection{Schramm's formula}\label{SubSecSchramm}
Schramm's formula relates to the following question:
given an SLE connecting two boundary points, what is the probability that
the curve passes to one side of a point, $\xi(z)$, in the
interior? If the domain is mapped to the upper half plane such
that the SLE connects $0$ to $\infty$, this probability may be
expressed as the solution to a second order partial differential
equation. Let us consider the probability that the curve passes to the
left of a point, $P_{\rm left}$. By this we mean that there is a
continuous path from the point to the positive real axis which
does not intersect the SLE curve. The solution is found by
applying appropriate boundary conditions to the solution to the
differential equation, which is a function only of $t\equiv
{\tr Re}(\xi)/{\tr Im}(\xi)$ due to the conformal invariance of the measure:

\begin{equation}\label{EqSchramm}
P_{\rm
left}(t)=\h-\frac{\g(\frac{4}{\kappa})}{\sqrt{\pi}\g(\frac{8-\kappa}{2\kappa})}t\,_{2}F_{1}
(\h,\frac{4}{\kappa};\frac{3}{2};-t^2) \,.
\end{equation}

The two domains considered in this work are related to the upper
half plane by simple transformations, hence this result may be
re-expressed easily in terms of the conventional coordinates in
these domains.

The generalisation of this result to the two-SLE case, was the
subject of a recent paper\cite{CardyGamsa}. Given a pair of SLEs
both connecting $0$ to $\infty$ in the upper half plane, there are
three possibilities: an interior point, $\xi(z)$, may be to the
left of both curves, between them or to the right hand side of
both. There is a probability for the curve to belong to each class, which is a
function only of $t\equiv Re(\xi)/Im(\xi)$, just as for the single
curve case. The probability of each class obeys a third
order linear partial differential equation, with solutions which
are integrals of hypergeometric functions. The particular solution
of interest is found by imposing a natural set of boundary
conditions, namely the asymptotic behaviour as the point
approaches the positive and negative real axis. The
solutions are

\begin{equation}\label{EqSchrammLeft}
\pl(t)=\frac{\Gamma(\frac{4}{\kappa})\Gamma(\frac{8}{\kappa})}
{2^{2-8/\kappa}\pi\Gamma(\frac{12}{\kappa}-1)}
\int_{t}^{\infty}S(t')\dd t'\,,
\end{equation}
where
\begin{align*}
S(t)=\frac{\,_{2}F_{1}(\frac{1}{2}+\frac{4}{\kappa},1-\frac{4}{\kappa};\frac{1}{2};-t^2)
-\frac{2\Gamma(1+\frac{4}{\kappa})\Gamma(\frac{4}{\kappa})}{\Gamma
(\frac{1}{2}+\frac{4}{\kappa})\Gamma(-\frac{1}{2}+\frac{4}{\kappa})}\,
t\,_{2}F_{1}(1+\frac{4}{\kappa},\frac{3}{2}-\frac{4}{\kappa};\frac{3}{2};-t^2)}{(1+t^2)^{\frac{8}{\kappa}-1}}\,,
\end{align*}
and
\begin{equation}\label{EqSchrammMid}
\pmm(t)=1-\frac{\Gamma(\frac{4}{\kappa})\Gamma(\frac{8}{\kappa})}
{2^{2-8/\kappa}\pi\Gamma(\frac{12}{\kappa}-1)}\Big[
\int_{t}^{\infty}S(t')dt'+\int_{-t}^{\infty}S(t')dt'\Big]\,.
\end{equation}
The probability that the curves are both to the right of a point may be deduced
from the other two cases.
Throughout this paper, we shall refer to these equations as the two-curve formulae.
As for the single curve
formula, the results may be transferred to the domains of interest
via a conformal transformation.

\subsection{4-leg operators}\label{Sec4leg}
Consider the $q=3$ Potts model with `fixed' boundary conditions.
There are a pair of cluster boundary curves in this case, which
initially overlap, taking the form of a `composite' cluster
boundary. This may be expected to split around a cluster of spins
which is not connected to the boundary. We may ask about the
relevance, in a renormalization group sense, of the operator
responsible for the recombination of these `split' curves. This
operator is familiar from the height representation of the $O(n)$
model, where it is known as the 4-leg operator. The scaling
dimension of this 4-leg operator is
\begin{equation}\label{Eq4legScaling}
  x_{4}=2g-\frac{(g-1)^2}{2g}\,
\end{equation}
where $g=1-\arccos{(\sqrt{q}/2)}/\pi$. For $q=4$, which we do not
consider in this paper, the 4-leg operator is therefore marginal
with $x_{4}=2$, since $g=1$ in this case. For $q=3$, the scaling
dimension is $x_{4}=99/60$, which means it is irrelevant. As the
system size increases, the `split' boundary curves are
therefore expected to collide less frequently. We may expect the scaling
limit in this case to correspond to a pair of non-intersecting SLE
curves.
Note that the joining of two `split' curves and a `composite' curve at a point
does not correspond to the insertion of a 3-leg operator, since the two types
of curves are inequivalent. Indeed, we shall see that the `composite' curve has
a different fractal dimension.

\section{Numerical simulations}\label{SecSimulations}
We use the algorithm described in appendix~\ref{SecWolffAlgorithm}
to generate a representative set of spin configurations at the
critical point of the $q$-state Potts model for both $q=2$ and $q=3$
in two domains. The first is the square lattice on the rectangle
with aspect ratio $3:1$, the second is the triangular lattice on
the disc.

As explained in section~\ref{SecPotts}, if the boundary conditions
for $q=2$ (the Ising model) are chosen appropriately, a continuous
boundary curve separating the two spin clusters runs between two
points on the boundary. This ensemble of curves has recently been
proven to converge to SLE$_{3}$~\cite{SmirnovIsing}. It is therefore a
useful check to show that the set of boundary curves obtained from
our algorithm for $q=2$ satisfy the expected properties of SLE, as
outlined in section~\ref{SecSLE}.

The main aim of this paper is to gather numerical support for the
spin boundaries of the 3-state Potts model with appropriate
boundary conditions becoming SLE in the scaling limit. There are two types of boundary
conditions: `fluctuating' boundary conditions (see
section~\ref{SecPotts}) lead to a single curve running between two
boundary points. `Fixed' boundary conditions lead to a pair
of `split' curves running between two fixed points on the
boundary, which may overlap to form a `composite' curve. In
section~\ref{Sec4leg} we presented an argument for only the
`split' curves being present in the scaling limit. We test the validity of this
argument by comparing the fractal dimensions of the two types of curves.

The geometries chosen for the simulations are the square lattice
on a narrow rectangular sample, and the triangular lattice on a
disc. The first is chosen so that Schramm's formula (see
section~\ref{SubSecSchramm}) may be compared with that on an
infinite strip. The second geometry is chosen for the ease of
obtaining the driving function of single curves (see
section~\ref{SecDrivingFunction}). The two are complimentary in
the sense that the measure on curves in the two domains should be
related by the conformal transformation between them. That
critical statistical models are conformally invariant is an
accepted conjecture, but this does provide another useful check of
the robustness of the algorithm, as well as a further
demonstration that properties of the scaling limit of the curves
are consistent with those of SLE.

The remainder of this section is dedicated to a technical
discussion of the methods of obtaining data from the samples.

\subsection{Identifying the cluster boundaries}\label{SecIdent}
\begin{figure}[htbp]
  \vspace{9pt}

  \centerline{\hbox{ \hspace{0.0in}
    \epsfxsize=1.6in
    \epsffile{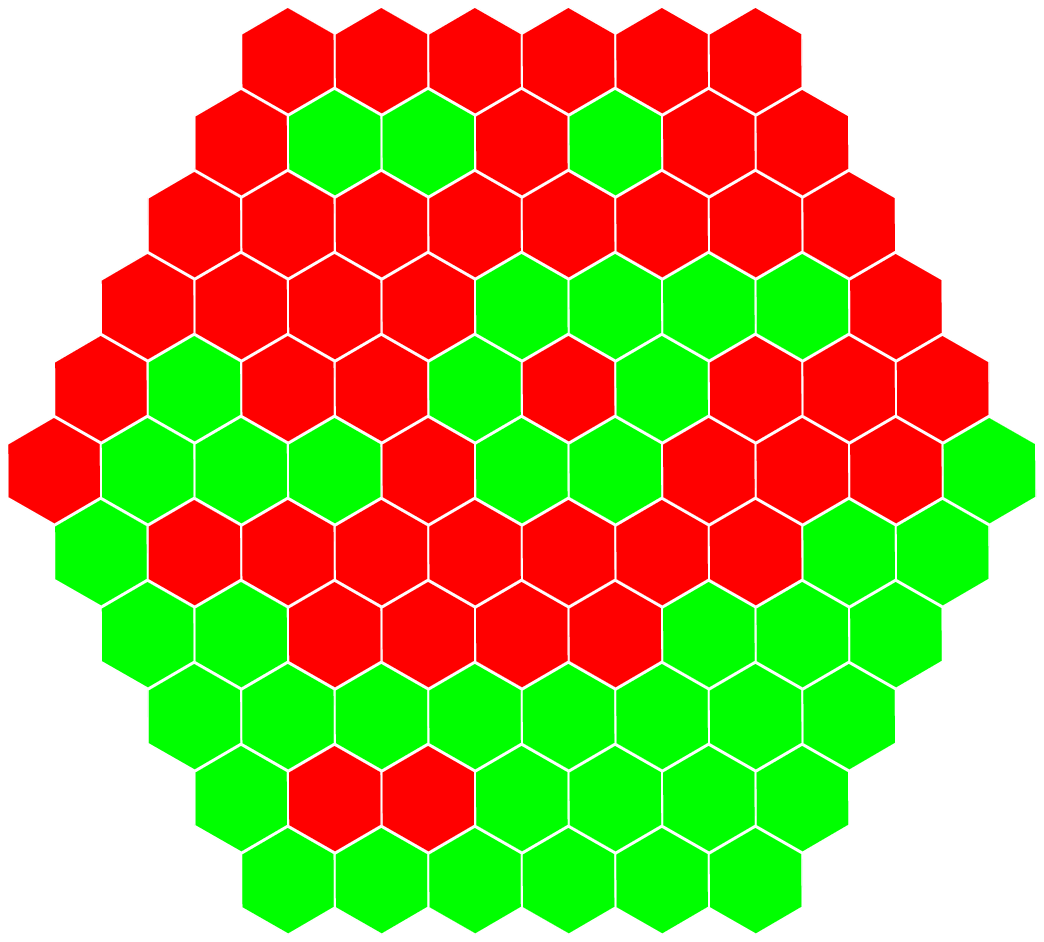}
    \hspace{0.25in}
    \epsfxsize=1.6in
    \epsffile{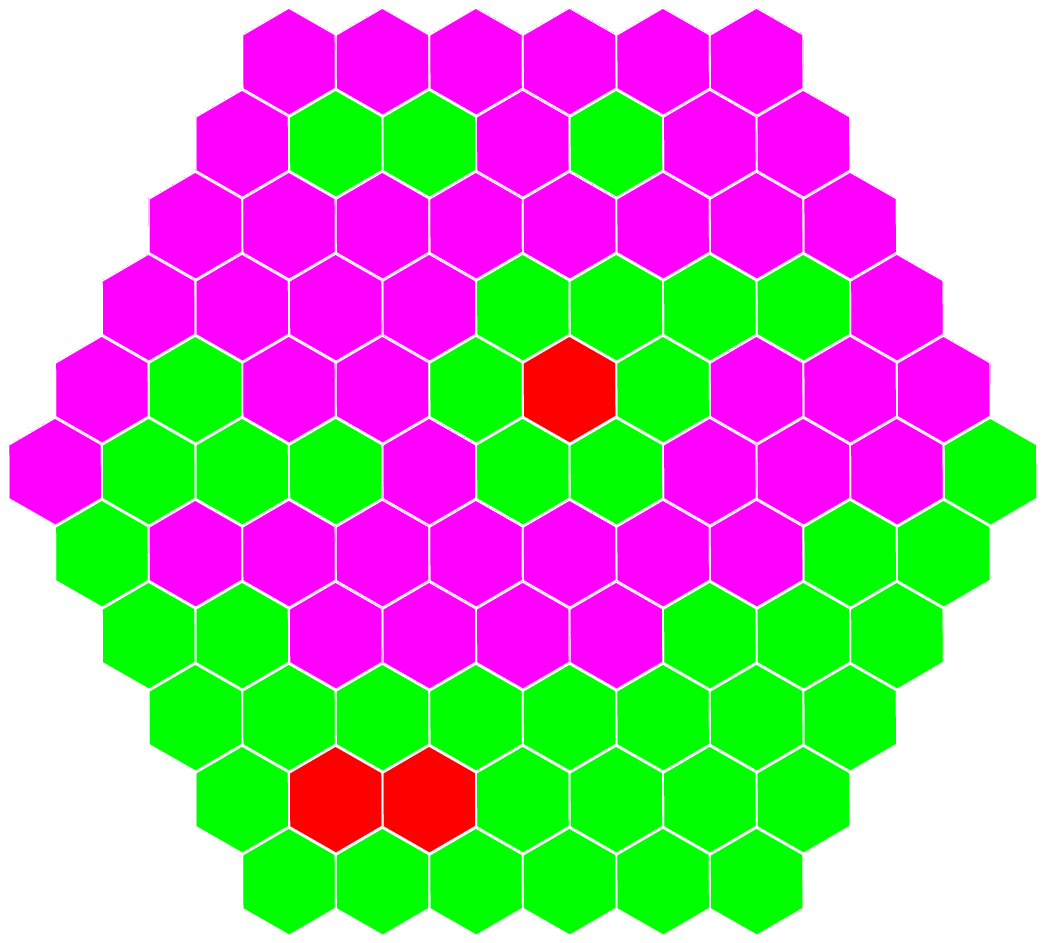}
    }
  }

  \vspace{4pt}
  \hbox{\hspace{1.3in} (a) \hspace{1.70in} (b)}
  \vspace{9pt}

  \centerline{\hbox{ \hspace{0.0in}
    \epsfxsize=1.6in
    \epsffile{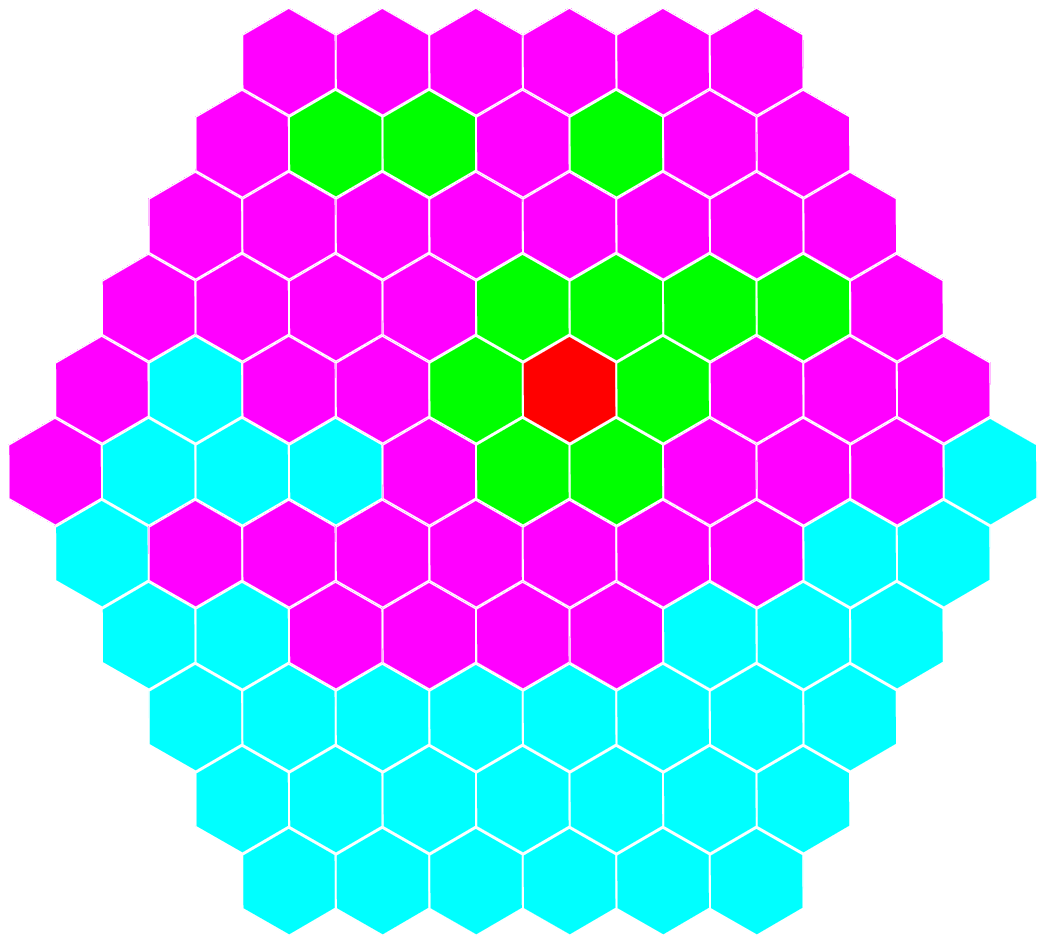}
    \hspace{0.25in}
    \epsfxsize=1.6in
    \epsffile{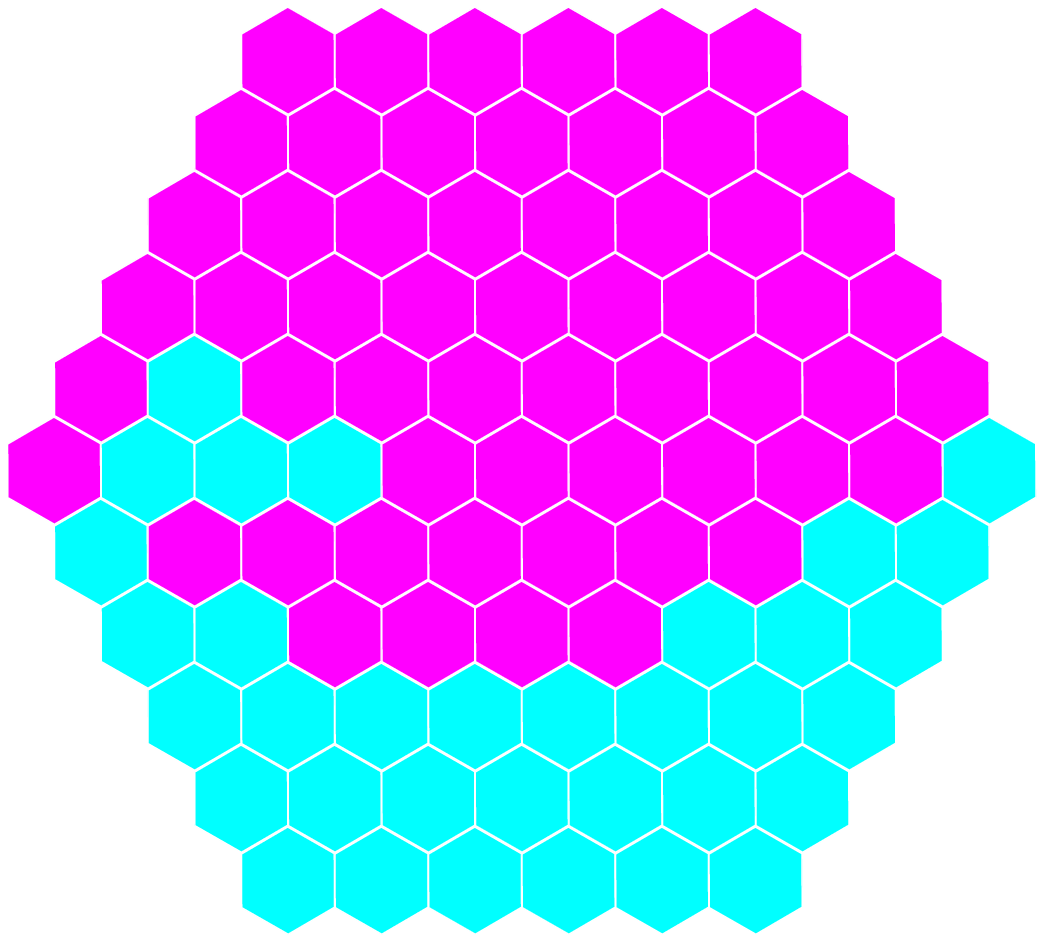}
    }
  }

  \vspace{4pt}
  \hbox{\hspace{1.3in} (c) \hspace{1.70in} (d)}
  \vspace{9pt}

  \caption{(a) shows a spin configuration in the Ising model. To obtain (b), recolour a
red boundary spin at the top of the sample purple. Recolour all
its red neighbours purple and repeat for all their neighbours et
cetera. To obtain (c), recolour a green spin on the bottom edge
cyan. Recolour all its non-purple neighbours cyan and repeat for
all their neighbours et cetera. Finally, to obtain (d), recolour
all remaining green and red spins purple.}
  \label{sub-figIs}

\end{figure}

Given a configuration of spins, the first step to identifying the
boundary curves is to identify the extent of the cluster
containing a given boundary spin. This is done using the following
algorithm:
\begin{enumerate}
  \item Start at the boundary spin of interest,
marking it as belonging to the cluster
  \item Mark all neighbours if they are of the same spin type
  \item Repeat step 2 for all neighbours of marked spins
  \item Identify all connected sets of unmarked spins which are
surrounded by a closed path of marked spins. Mark all such sets of
spins.
\end{enumerate}
This process is demonstrated in figure~\ref{sub-figIs} for an
Ising sample.

It is sufficient to identify a single cluster boundary for the
Ising model and $q=3$ model with `fluctuating' boundary
conditions, since the cluster boundaries of the top and bottom
clusters overlap everywhere in these cases (see
section~\ref{SecPotts}). For the case $q=3$ with `fluctuating'
boundary conditions, as in figure~\ref{Figq3Bound} for example,
the first step to identifying the mixed spin cluster boundary is
to replace all blue spins throughout the sample with green spins.
Then, the algorithm is the same as that for the Ising model.

\begin{figure}[htbp]
  \vspace{9pt}

  \centerline{\hbox{ \hspace{0.0in}
    \epsfxsize=1.4in
    \epsffile{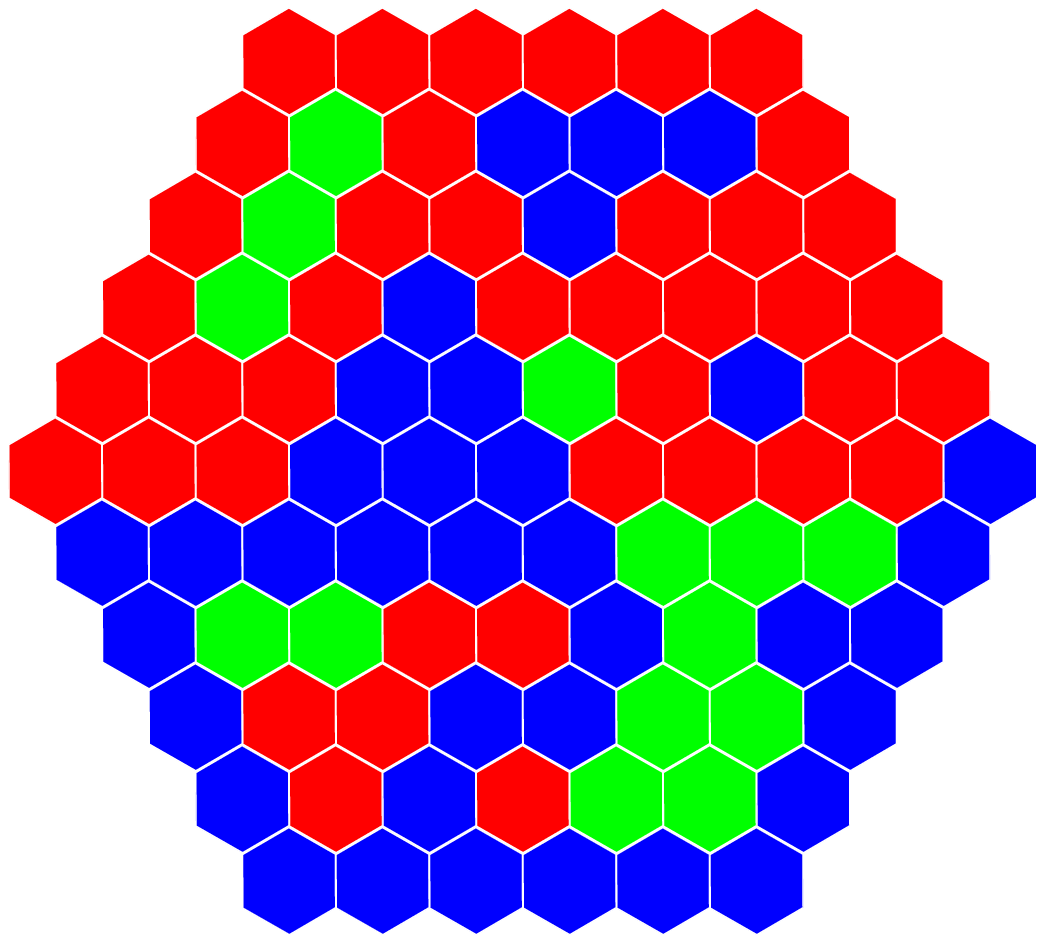}
    \hspace{0.1in}
    \epsfxsize=1.4in
    \epsffile{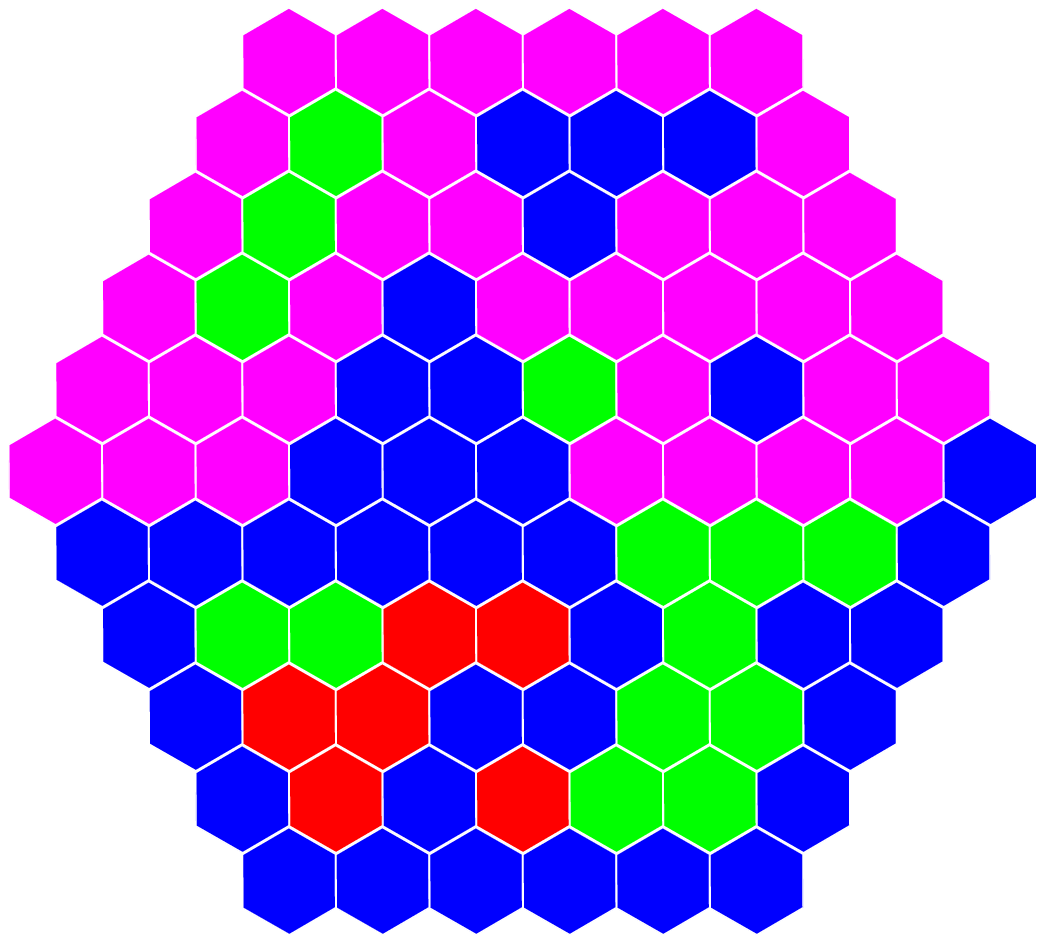}
    \hspace{0.1in}
    \epsfxsize=1.4in
    \epsffile{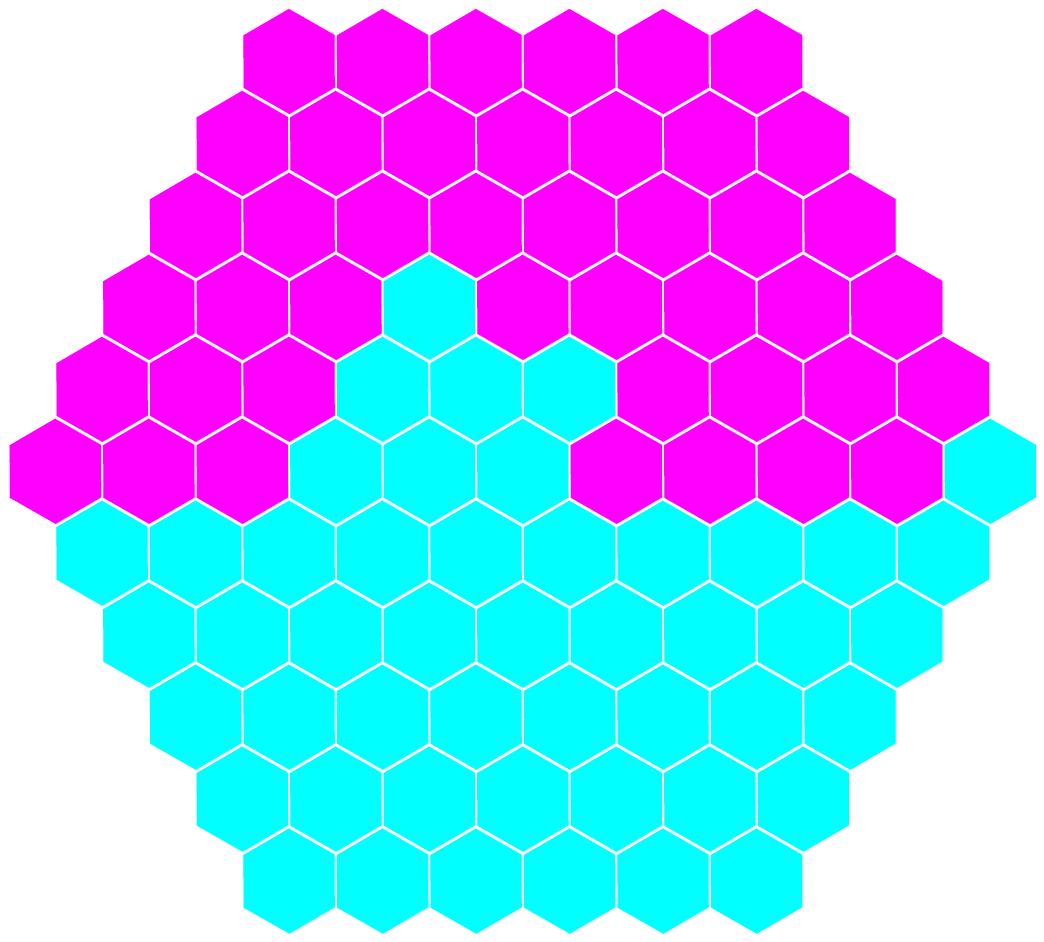}
    }
  }

  \vspace{4pt}
  \hbox{\hspace{.7in} (a) \hspace{1.3in} (b) \hspace{1.3in} (c)}
  \vspace{9pt}

  \centerline{\hbox{ \hspace{0.0in}
    \epsfxsize=1.4in
    \epsffile{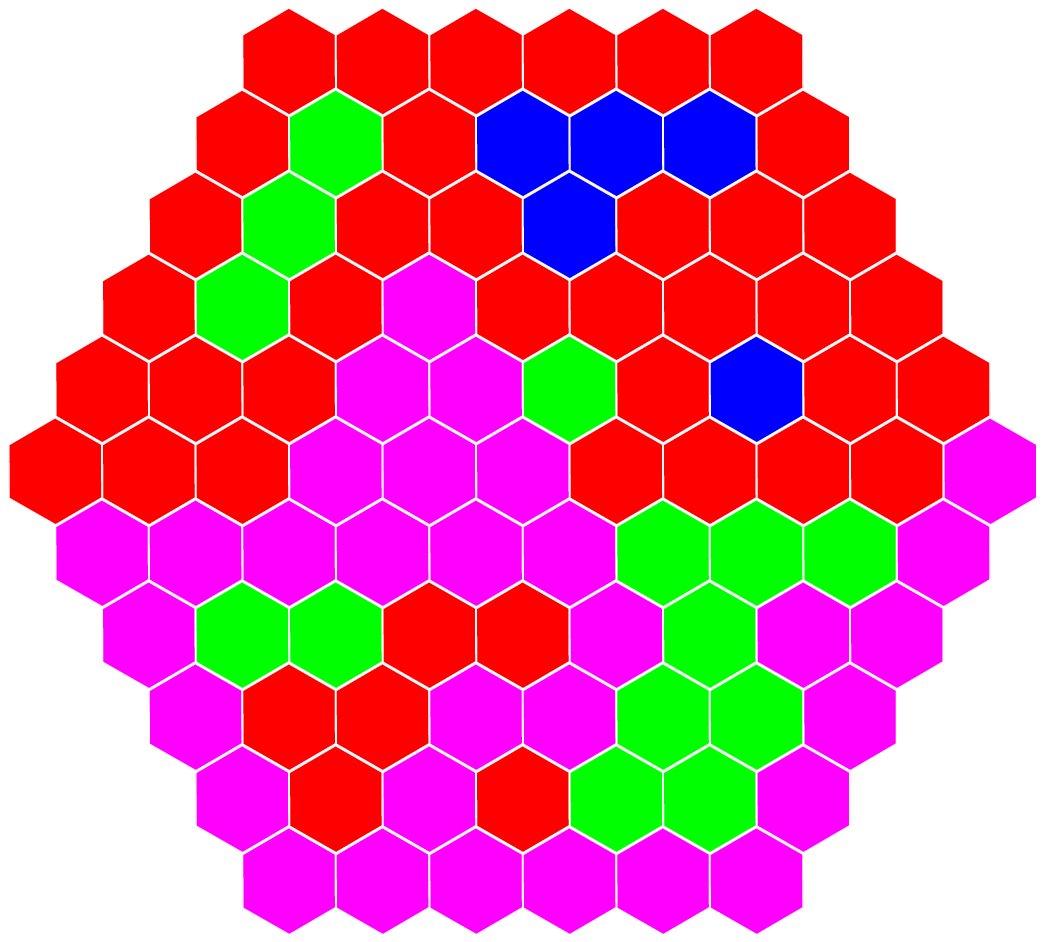}
    \hspace{0.1in}
    \epsfxsize=1.4in
    \epsffile{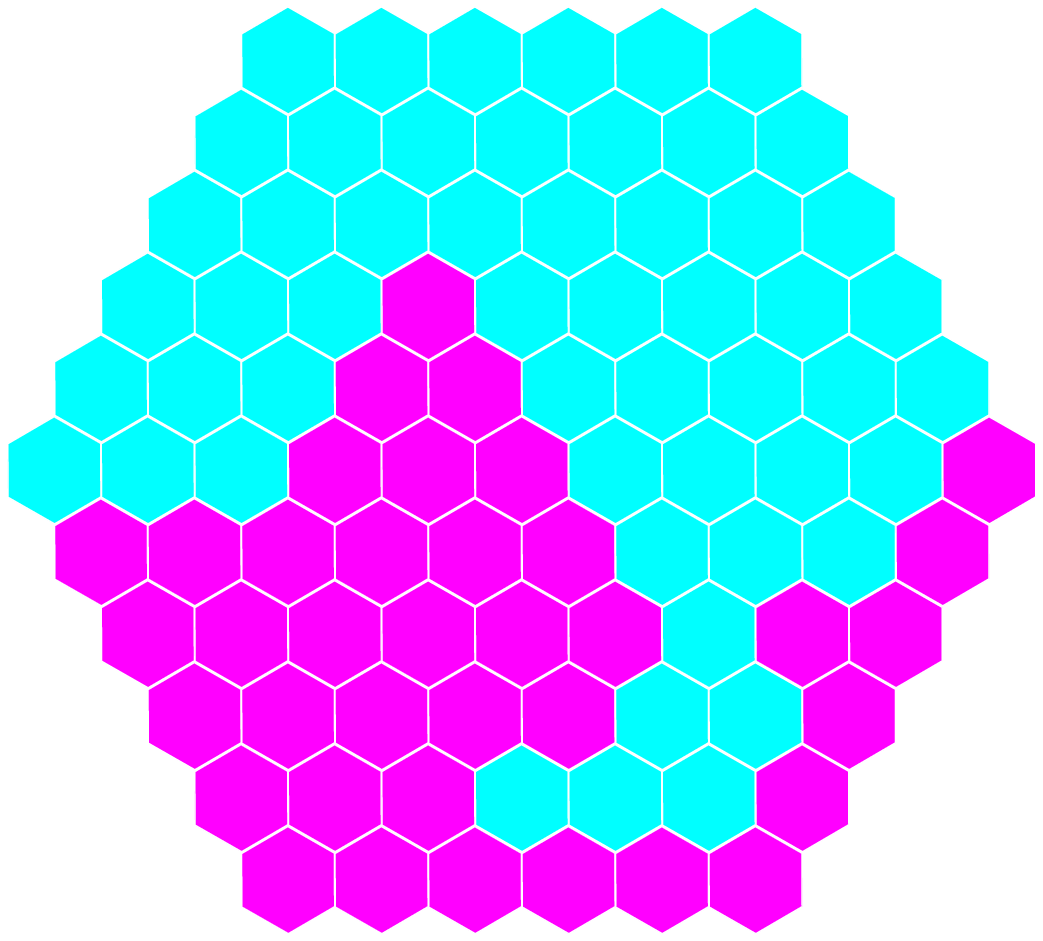}
    \hspace{0.1in}
    \epsfxsize=1.4in
    \epsffile{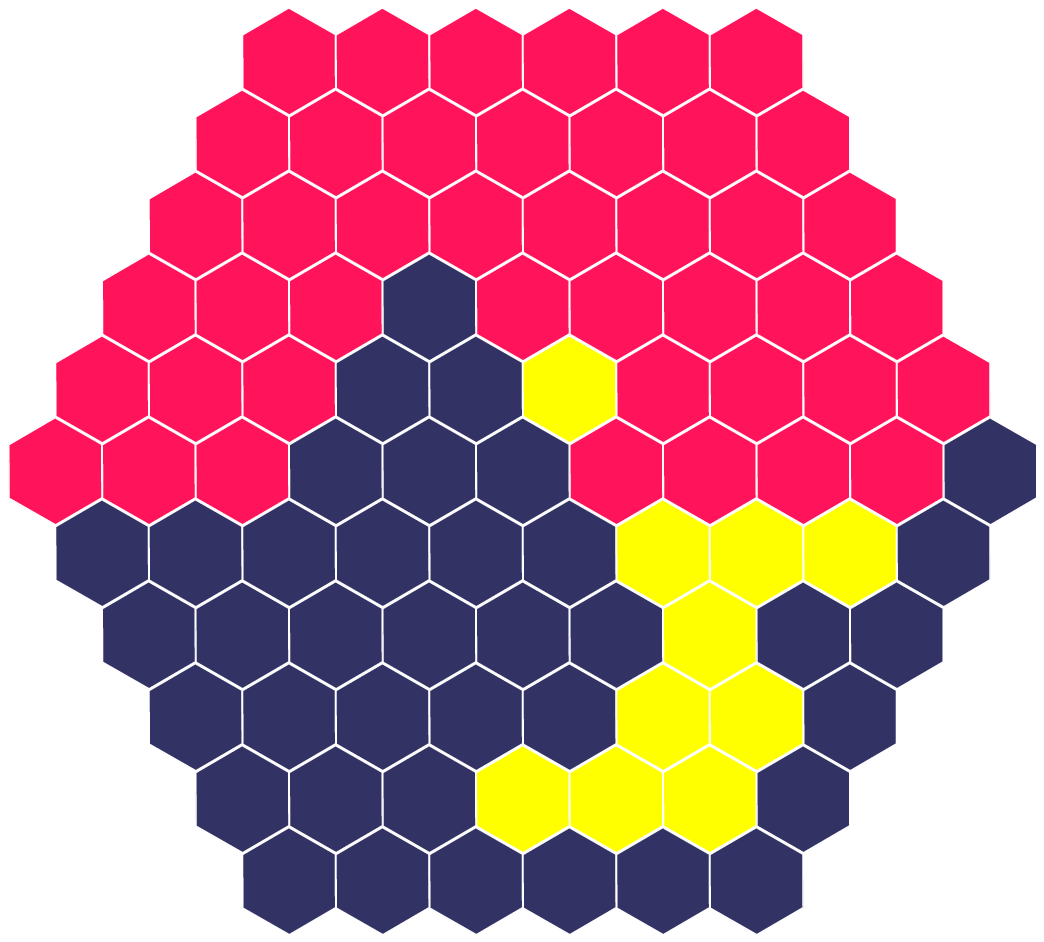}
    }
  }

  \vspace{4pt}
  \hbox{\hspace{.7in} (d) \hspace{1.3in} (e) \hspace{1.3in} (f)}
  \vspace{9pt}

  \caption{Figure (a) shows a $q=3$ Potts model spin configuration with `fixed' boundary conditions.
To obtain (b), recolour a red boundary spin at the top of the
sample purple. Recolour all its red neighbours purple and their
red neighbours purple et cetera. To obtain (c), recolour a blue
spin on the bottom edge cyan and all its non-purple neighbours
also. Repeat for all their non-purple neighbours et cetera.
Recolour purple all the remaining red, blue and green spins. To
identify the cluster containing the blue boundary spins, start
again from figure (a). Fig (d) shows the result of starting with a
blue spin on the bottom edge and recolouring all its blue
neighbours purple, et cetera. Fig (e) is obtained after
recolouring cyan all non-purple neighbours starting from a red
spin at the top edge. Lastly, recolour all remaining red, blue and
green spins purple. The two clusters connected to the boundary are
coloured differently in figures (c) and (e). But some spins are
cyan in both (c) and (e). These are yellow in figure (f). The top
cluster is pink and the bottom cluster is grey.}
  \label{sub-figq3}

\end{figure}

For the $q=3$ model with `fixed' boundary conditions, however,
this process must be run twice to identify both the cluster
connected to the bottom edge of the sample and that connected to
the top edge, see figure~\ref{sub-figq3}. The result is
figure~\ref{sub-figq3}(f), which shows all three types of clusters
present. Note that the example figures are for the triangular lattice case. The same
algorithms to determine the cluster boundaries are employed for samples
generated on the square lattice.

\subsection{Estimating the fractal dimension}\label{SecNumFract}
For the Ising model and the $q=3$ Potts model with `fluctuating'
boundary conditions, there is a single curve running between two
fixed points on the boundary. First, identify the cluster
containing one set of boundary spins, as outlined in
section~\ref{SecIdent}. This results in a configuration of the
type shown in figure~\ref{sub-figIs}(d). Then, the length of the
cluster boundary curve is deduced as follows. For each site on the
sample, compare the colour at that site to each of its nearest
neighbours. If the nearest neighbour is of the same colour (and
therefore connected to the same cluster) there is not a cluster boundary
separating the two spins. Otherwise, if it is
of the other colour, add one half to the total length of the
cluster boundary curve (since each link will be counted twice by
this algorithm).

The situation for $q=3$ with `fixed' boundary conditions involves
the fractal dimension of two types of cluster boundaries: the
`composite' cluster boundaries and the `split' cluster boundaries,
see section~\ref{SecPotts}. Identify the clusters containing the
two types of boundary spin and also the clusters which do not
contain a boundary spin. This leads to a cluster configuration
similar to that shown in figure~\ref{sub-figq3}(f). The total
length of `composite' boundaries is found by summing over all
sites, adding one half to a running total wherever a pink site is
adjacent to a grey site. The total length of `split' cluster
boundaries is found by identifying all sites where a grey site is
adjacent to a yellow site or a pink site is adjacent to a yellow
site, adding one half to a different running total for each such
occurence.

In this way, we may deduce the expected length of curves as a
function of system size. Comparison to equation~\ref{EqFractLength}
allows an estimation of the fractal dimension of the curves.

\subsection{Comparison to Schramm's formula and its generalisation to two curves}\label{SecNumCross}
Firstly, for the Ising model and $q=3$ Potts model with
`fluctuating' boundary conditions, the starting point is the
cluster configuration picture, as in figure~\ref{sub-figIs}(d).
Recall that the boundary conditions are chosen such that the curve
starts and ends at opposite points on the boundary. There is a
path which is equidistant from these two points, which is
shown in figure~\ref{FigIsRectGeom} for the rectangle. For each site along this
path, calculate the percentage of the samples for which the site
is coloured cyan. This yields the fraction of samples for which
this point is connected to the bottom cluster. This may be
compared to the predictions of Schramm's formula.

For the $q=3$ Potts model with `fixed' boundary conditions, the
procedure is to deduce the cluster picture, as in
figure~\ref{sub-figq3}(f). Two separate quantities may be
measured, the percentage of the samples for which each point along
the path is connected to the bottom cluster (coloured grey) and
the percentage which are in the central, yellow cluster. In the
former case, the point is to one side of both curves (the left
hand side say), whereas in the latter example, the points are
between the two curves. These results are to be compared to the
generalisation of Schramm's formula to two curves, the two-curve formulae (see
section~\ref{SubSecSchramm}).

\subsection{The driving function}\label{SecDrivingFunction}
The third test is to analyse the driving function of the curves
for the Ising model and $q=3$ Potts model samples with
`fluctuating' boundary conditions. The spin boundary curves have a
natural discretisation provided by the lattice itself, so a curve
may be approximated by the set of ordered lattice points which it
passes through. It is easiest to consider the set of conformal
mappings of the curve back to the boundary in the upper half
plane, so the first step is to transform domains using a conformal
transformation. For the disc, for example, this is the following
Mobius transformation:
\begin{equation}\label{EqMapdiscUHP}
h(z)=\frac{z-1}{i(z+1)}\,.
\end{equation}
For the rectangle, the transformation to the upper half plane is
not so convenient; the derivative of the conformal map at the
corners is large. The effect of this cannot be simply and
systematically removed.

The curve grown up to a point $\gamma$ may be mapped back to the
real axis by a series of vertical slit maps for each successive
discretized point, $\chi_{i}$, along the curve
\begin{equation}\label{EqSlitMap}
  g_{i,t}(z)=\re(\chi_{i})+\sqrt{((z-\re(\chi_{i}))^2+\im(\chi_{i})^2)}\,.
\end{equation}
This maps the point $\chi_{i}$ to $\re(\chi_{i})$ on the real
axis, which is therefore the value of the driving function,
$a_{t}$, at the time when this point is the tip of the growing
curve. The change in $t$ associated with this piece of curve is
found from the expansion of the map as $z\rightarrow\infty$:
\begin{equation*}
  g_{t}(z\rightarrow\infty)=z+\frac{\im(\chi)^2/2}{z}+\ldots
\end{equation*}
The change in $t$ is therefore $\delta t=\im(\chi_{i})^2/4$. For
SLE, the driving function should be a multiple of Brownian motion.
In particular, the expectation value of $a_{t}$ should vanish and
the variance should be $\kappa$. On a plot of frequency of step
lengths (normalised by $\sqrt{\delta t}$) the distribution should
be Gaussian with variance $\kappa t$. Our lattice curves are only
supposed to approximate SLE at scales much larger than the lattice
spacing. Therefore, effective steps equivalent to $10$ lattice steps
are considered. $\delta B$ is then the change in the driving
function as a result of an effective step, and $\delta t$ is the
capacity change, totalled over the $10$ component steps. We also
plot $\la a_{t}^2\ra$ as a function of $t$, with a view to
equating the gradient to $\kappa$.

\section{Results}\label{SecResults}
In this section, we present a comparison between the results from
the numerical simulations and predictions from SLE. The algorithm
used to generate samples is described in
appendix~\ref{SecWolffAlgorithm}.

\subsection{Ising model on the rectangle}\label{SubSecIsingRect}
\begin{figure}[thbp]
  \begin{center}
  \scalebox{0.4}{\centerline{
    \epsfig{figure=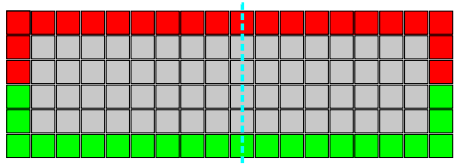}}}
    \caption{The geometry used for the Ising model on the rectangle of aspect ratio $1:3$
with a square lattice. The boundary conditions are shown, but the
bulk spins left grey. The path used to test against Schramm's
formula is shown as the blue dashed line.}
    \label{FigIsRectGeom}
  \end{center}
\end{figure}

\begin{figure}[thbp]
  \begin{center}
  \scalebox{0.35}{\centerline{
    \epsfig{figure=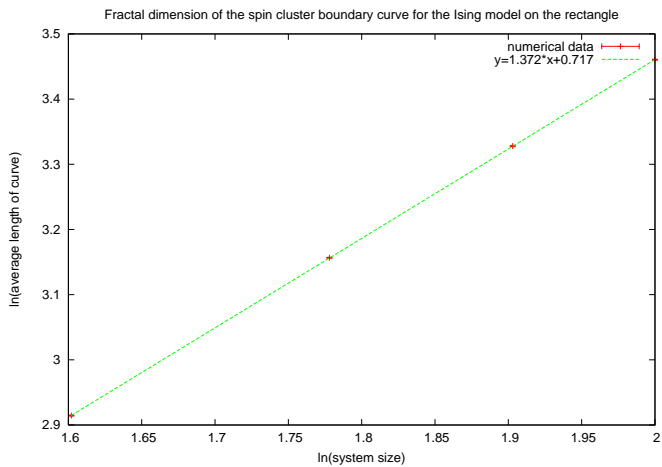, angle=270}}}
    \caption{The fractal dimension of the Ising model on the rectangle. The system size is
defined as the number of lattice spacings across the smaller side
of the rectangle. The curve length is measured in multiples of
lattice spacings.}
    \label{FigIsingRectFract}
  \end{center}
\end{figure}

This section describes the results from the Ising model on the
rectangle on a square lattice of aspect ratio $1:3$. Four system
sizes are used: $40\times 120$, $60\times 180$, $80\times 240$ and
$100\times 300$, where the lengths are in multiples of the lattice
spacing. These will be referred to by their smaller dimension. The
boundary conditions are as described in section~\ref{SecPotts},
with one spin type fixed along a long edge and half each of the
shorter sides, and the other spin type elsewhere. The geometry is
therefore as shown in figure~\ref{FigIsRectGeom}.
A cluster boundary curve runs from the centre of one short edge of
the rectangle to the centre of the other short edge. $50,000$
independent samples of each system size are used to estimate  the
fractal dimension of the curves, by determining the average length
as a function of system size. This process is explained in
section~\ref{SecNumFract}. The results are plotted in
figure~\ref{FigIsingRectFract}. The error bars are shown, but are
only just discernible at this magnification.
The linear fit is in good agreement with the data, with reduced
$\chi^2$ equal to $0.7$. It shows a fractal dimension of $d_{f}=1.372\pm
0.002$. Using formula~\ref{EqFract}, this would correspond to SLE
with $\kappa=2.976\pm 0.02$. The known value
of $\kappa=3$ for the Ising model lies just outside these error bars.

\begin{figure}[thbp]
  \begin{center}
  \scalebox{0.35}{\epsfig{figure=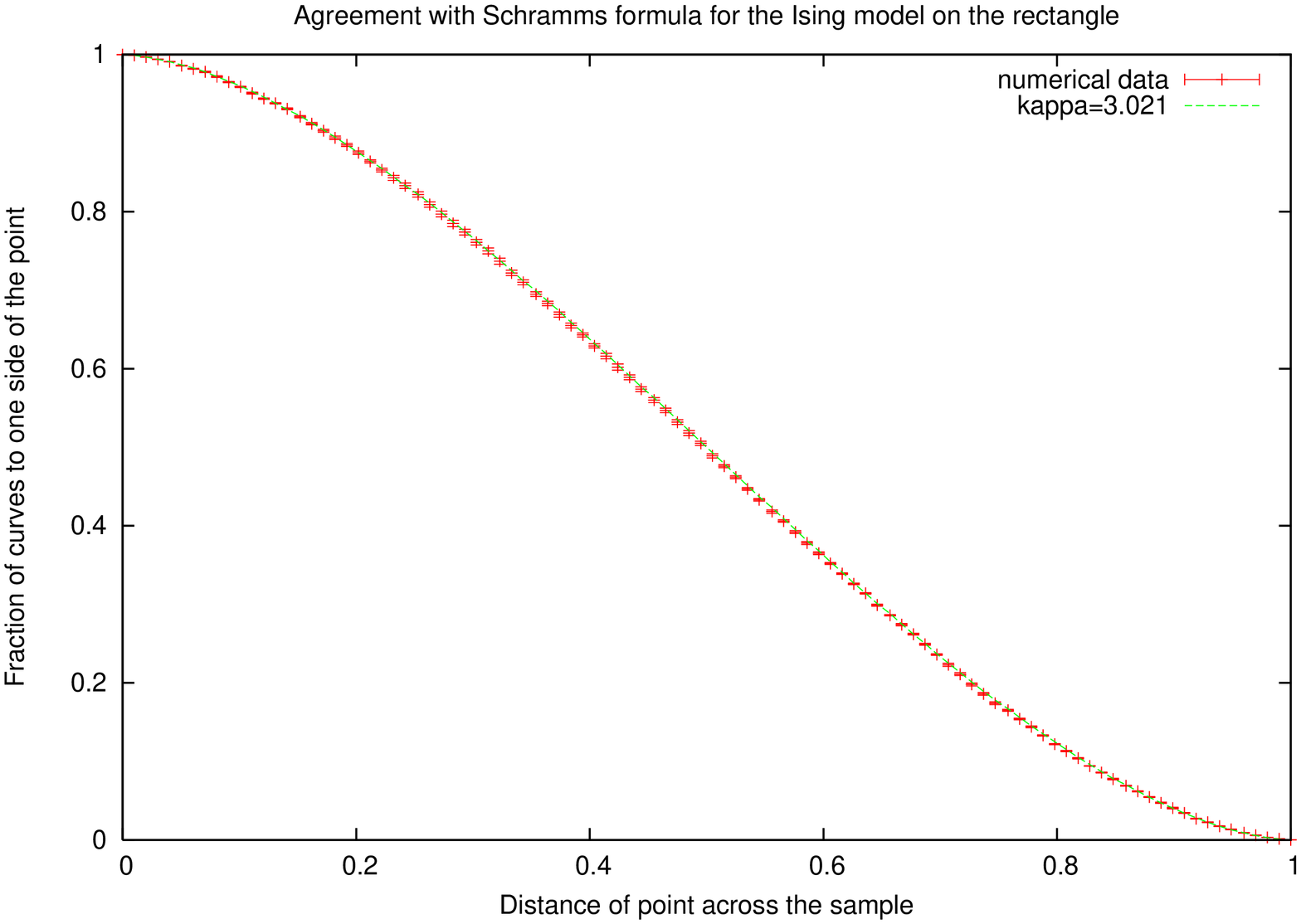, angle=270}}
  \scalebox{0.35}{\epsfig{figure=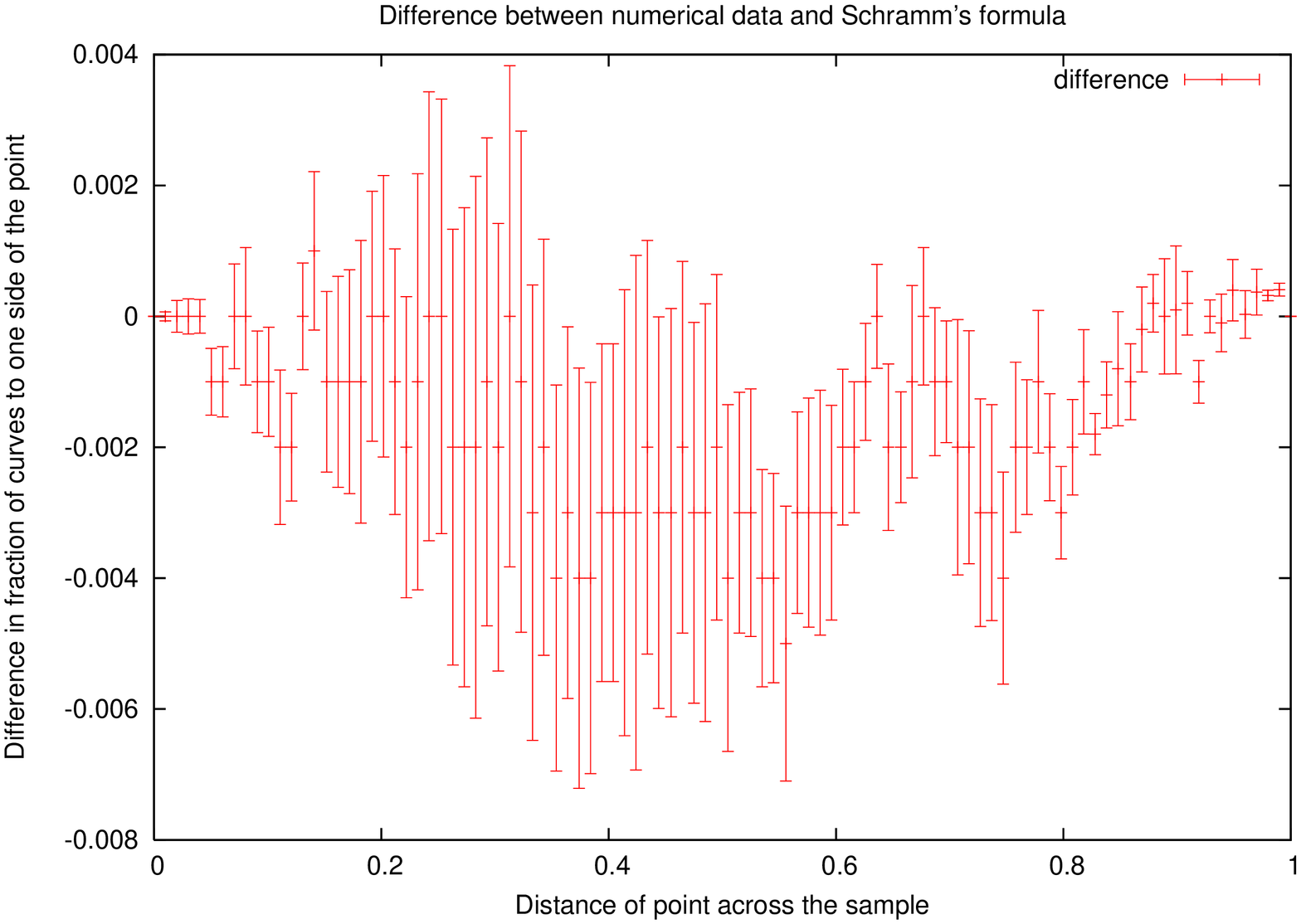, angle=270}}
    \caption{Fit to Schramm's formula on the strip for the Ising model,
system size $100\times 300$. The smaller system sizes also show
good agreement with Schramm's formula with $\kappa\approx
3.0$. The agreement is to approximately 1\%, as shown in the lower figure.}
    \label{FigSchrammIsingRect100}
  \end{center}
\end{figure}

The fit to Schramm's formula is via the logarithmic map
from the upper half plane to the infinite strip:
\begin{equation}\label{EqUHPtoPlane}
  w(z)=\frac{\ln{z}}{\pi}\,.
\end{equation}
The assumption that the strip is effectively infinitely long is
valid only if the boundary curve at the centre of the sample is
independent of the boundary conditions at the ends of the
rectangle. We tested this by generating system sizes of aspect
ratios $1:5$ and $1:6$. The differences in the
data are smaller than the error in the measurements. In terms of a
coordinate, $s$, which describes the fractional distance across
the strip, $t$ is given by $t=\cot(\pi s)$ in equation~\ref{EqSchramm}. This
follows from equation~\ref{EqUHPtoPlane}. The results from the
numerical data are the expectation value for the point at a
fractional distance $s$ across the strip being connected to the
bottom spin cluster. Figure~\ref{FigSchrammIsingRect100} shows the
comparison to the best fit value of $\kappa=3.021\pm 0.015$.
Once more, the known value lies just outside this range.
We should not be surprised, therefore, it the predicted value of $10/3$
for $q=3$ Potts model also lies just outside the range obtained from the numerics.
The average is again over $50,000$ independent samples.
\subsection{$q=3$ on the rectangle}\label{SubSecq3Rect}
\begin{figure}[thbp]
  \begin{center}
  \scalebox{0.35}{\centerline{
    \epsfig{figure=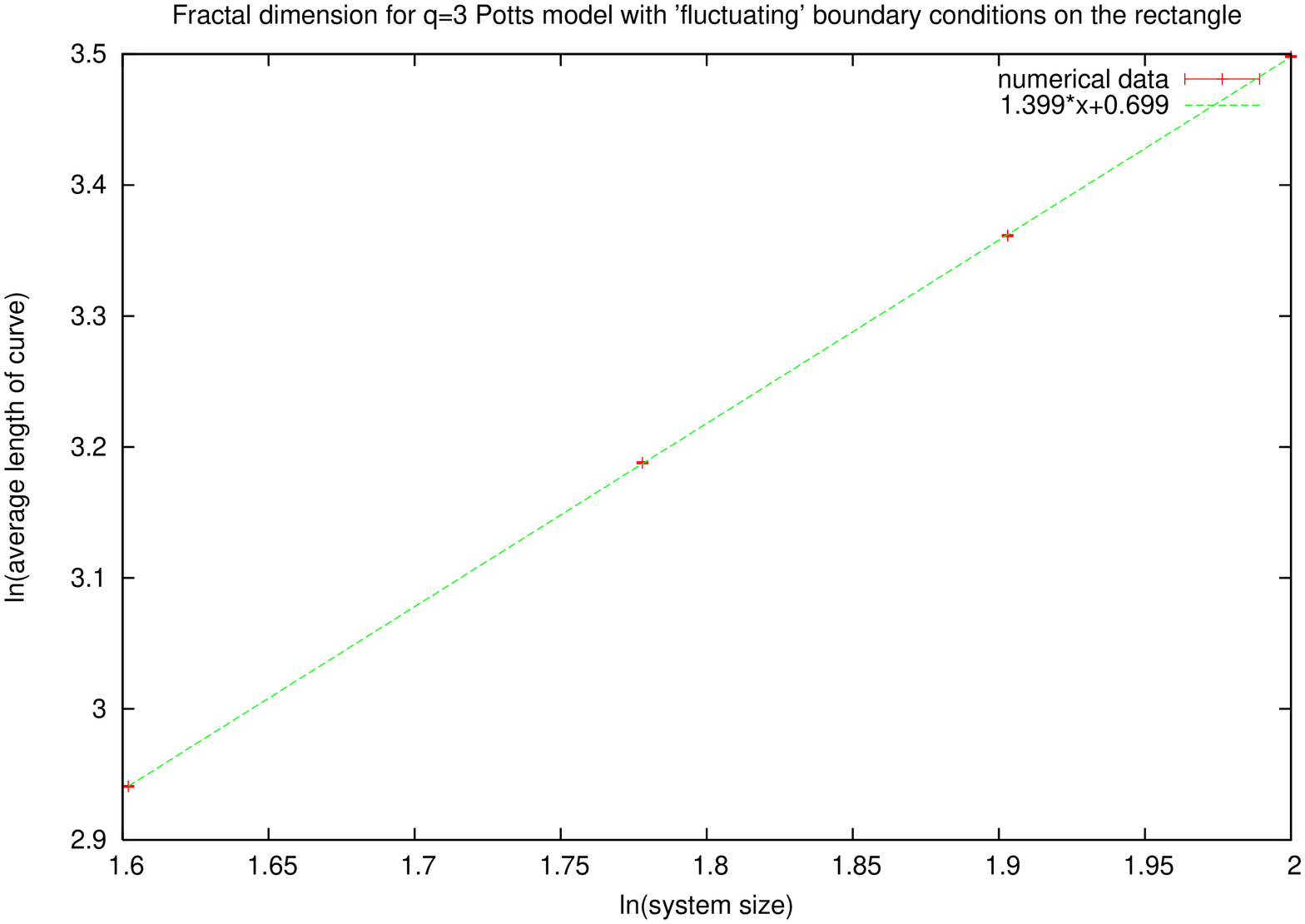, angle=270}}}
    \caption{The fractal dimension of the $q=3$ Potts model on the rectangle with
fluctuating boundary conditions. System size is the shorter side
of the rectangle in units of lattice spacing. The curve length is
in units of lattice spacings.}
    \label{Fig3qRectFract}
  \end{center}
\end{figure}

The geometry is the same as for the Ising model above, with sample
sizes $40\times 120$, $60\times 180$, $80\times 240$ and
$100\times 300$. First, let us examine the case of `fluctuating'
boundary conditions, when there is a single curve connecting two
boundary points at the centre of each short end of the rectangle.
The boundary conditions are like those in
figure~\ref{FigIsRectGeom} except that the boundary spins on the
bottom half are permitted to be either blue or green and to
fluctuate with the spins in the bulk. The fractal dimension comes
from the average length of the curves as a function of system
size, where $50,000$ independent samples have been generated for
each sample size. The results are shown in
figure~\ref{Fig3qRectFract}.
The linear fit shows the fractal dimension to be $d_{f}=1.399\pm
0.002$, which would correspond to SLE with $\kappa=3.192\pm 0.02$.
\begin{figure}[thbp]
  \begin{center}
  \scalebox{0.35}{\epsfig{figure=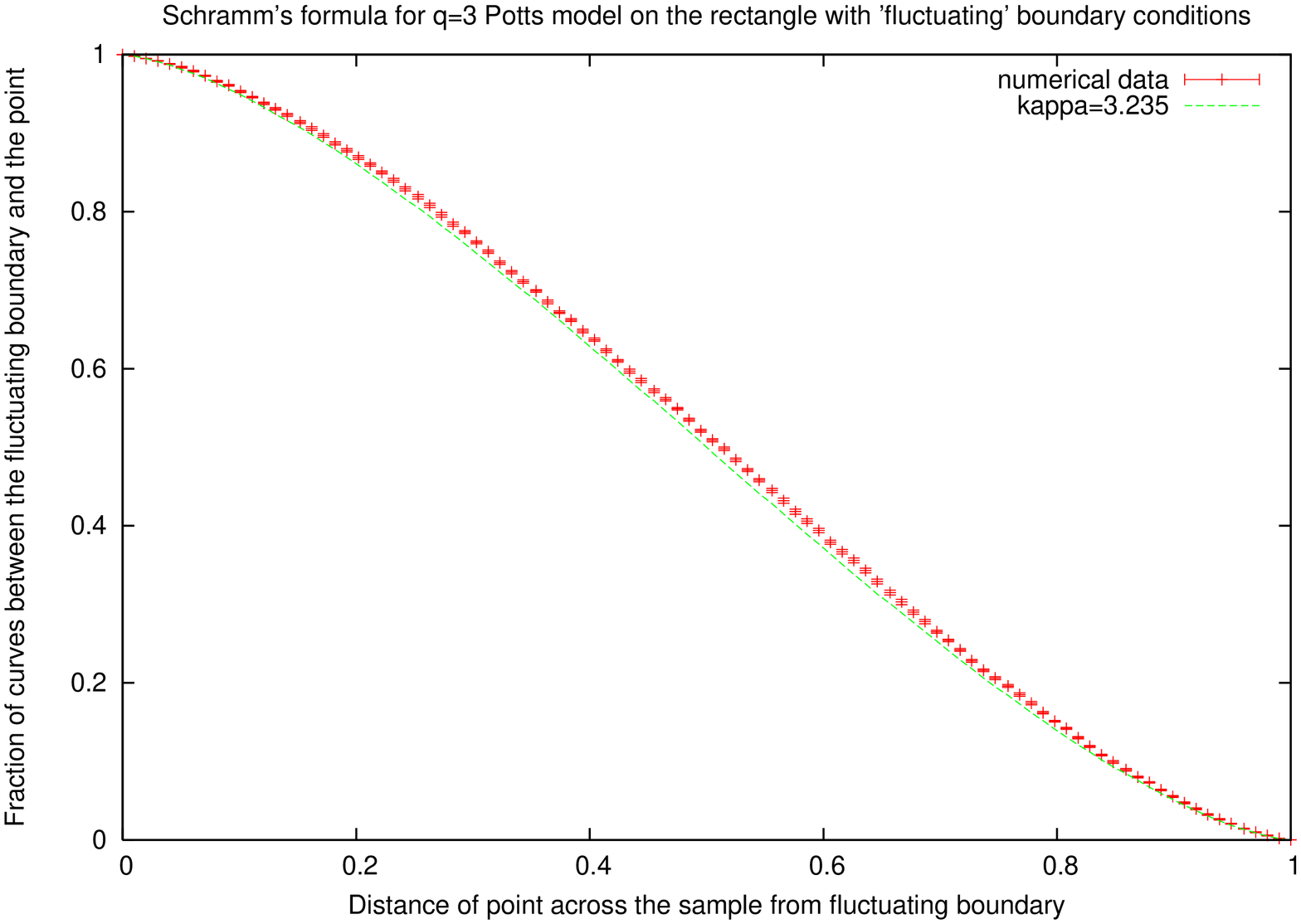, angle=270}}
  \scalebox{0.35}{\epsfig{figure=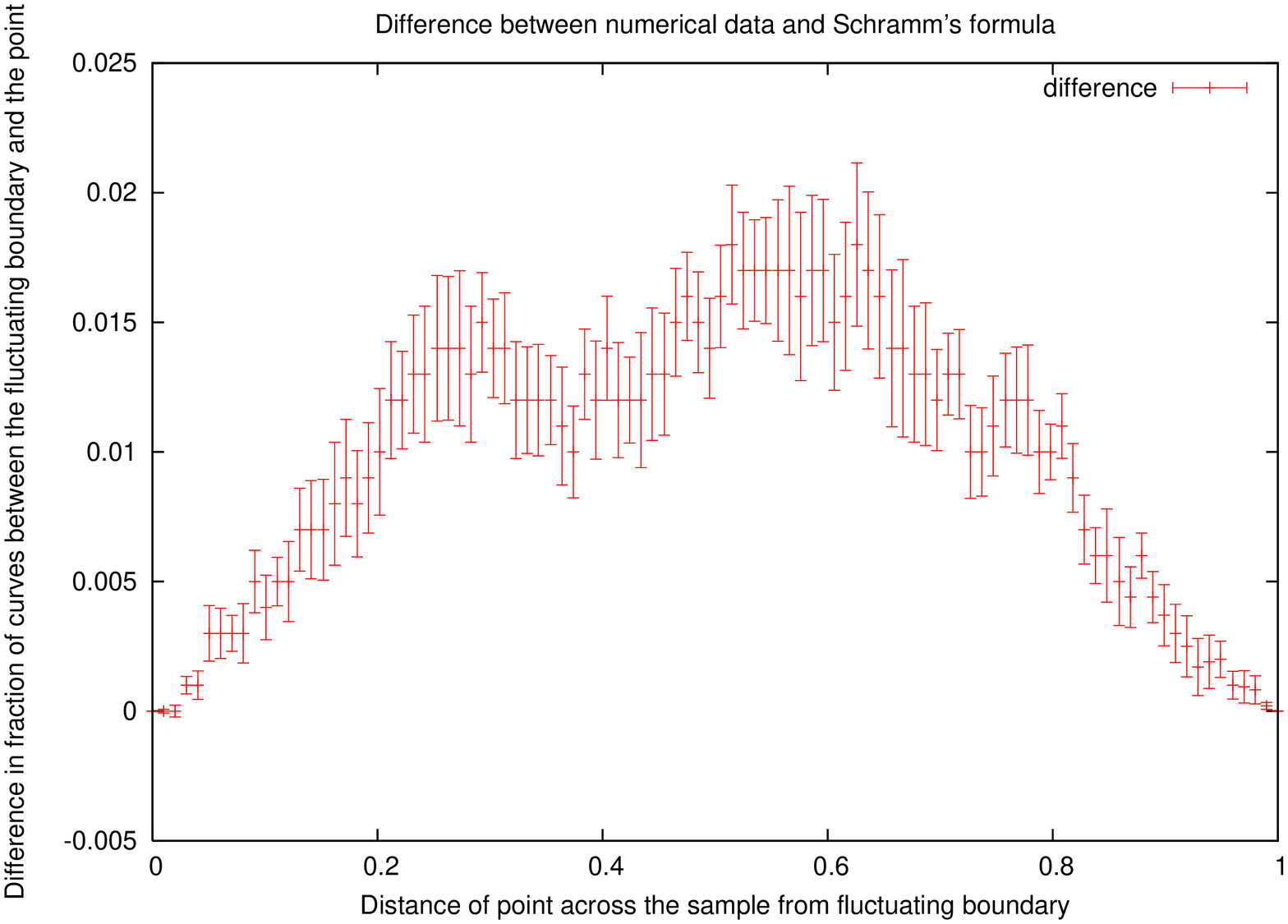, angle=270}}
    \caption{Schramm's formula for $q=3$ on the rectangle for system size $100\times 300$.
The distance across the sample is from the fluctuating boundary
toward the spin~$1$ boundary.}
    \label{Fig3qRectSchramm}
  \end{center}
\end{figure}
The fit to Schramm's formula,
figure~\ref{Fig3qRectSchramm}, shows that on average, the curve is
closer to the fluctuating spins boundary than would be expected.
Equivalently, a point of fractional distance $0\leq s\leq 1$
across the sample from the fluctuating boundary towards the fixed
boundary is more likely to be connected to the fixed spin cluster
than would be expected from Schramm's formula. Perhaps this is not
surprising; whereas the two types of boundary conditions in the
Ising model are reflection symmetric, the boundary conditions for
$q=3$ with fluctuating boundary conditions are not. It is only in
the scaling limit that we might expect to recover reflection
symmetry. The best fit value is $\kappa=3.235\pm 0.01$. We shall
see that this asymmetry is a result of non-zero average of the
driving function, but that this effect is localised to the
neighbourhood of the boundary spins, and therefore is
unimportant in the scaling limit.

\begin{figure}[thbp]
  \begin{center}
  \scalebox{0.35}{\epsfig{figure=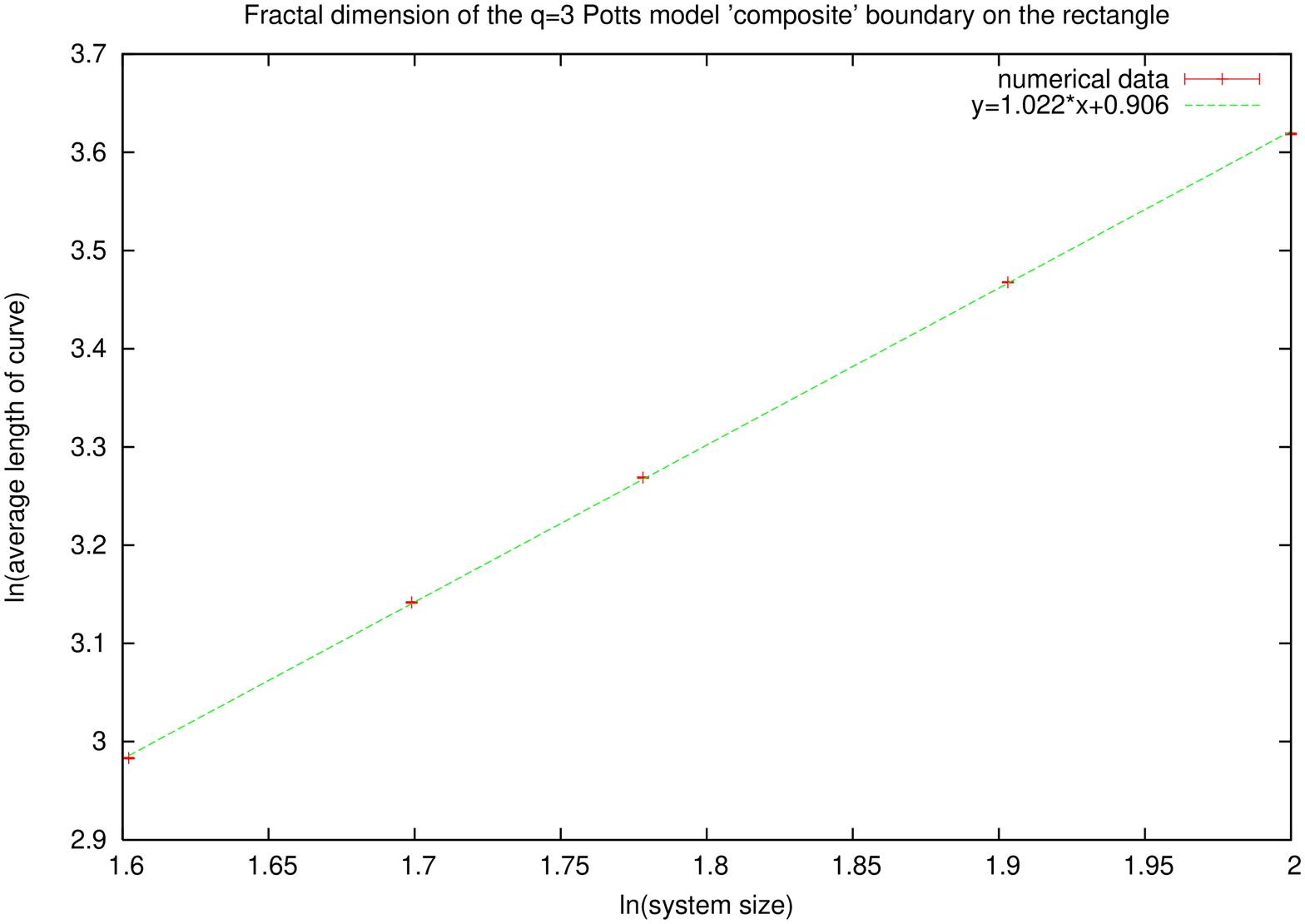, angle=270}}
  \scalebox{0.35}{\epsfig{figure=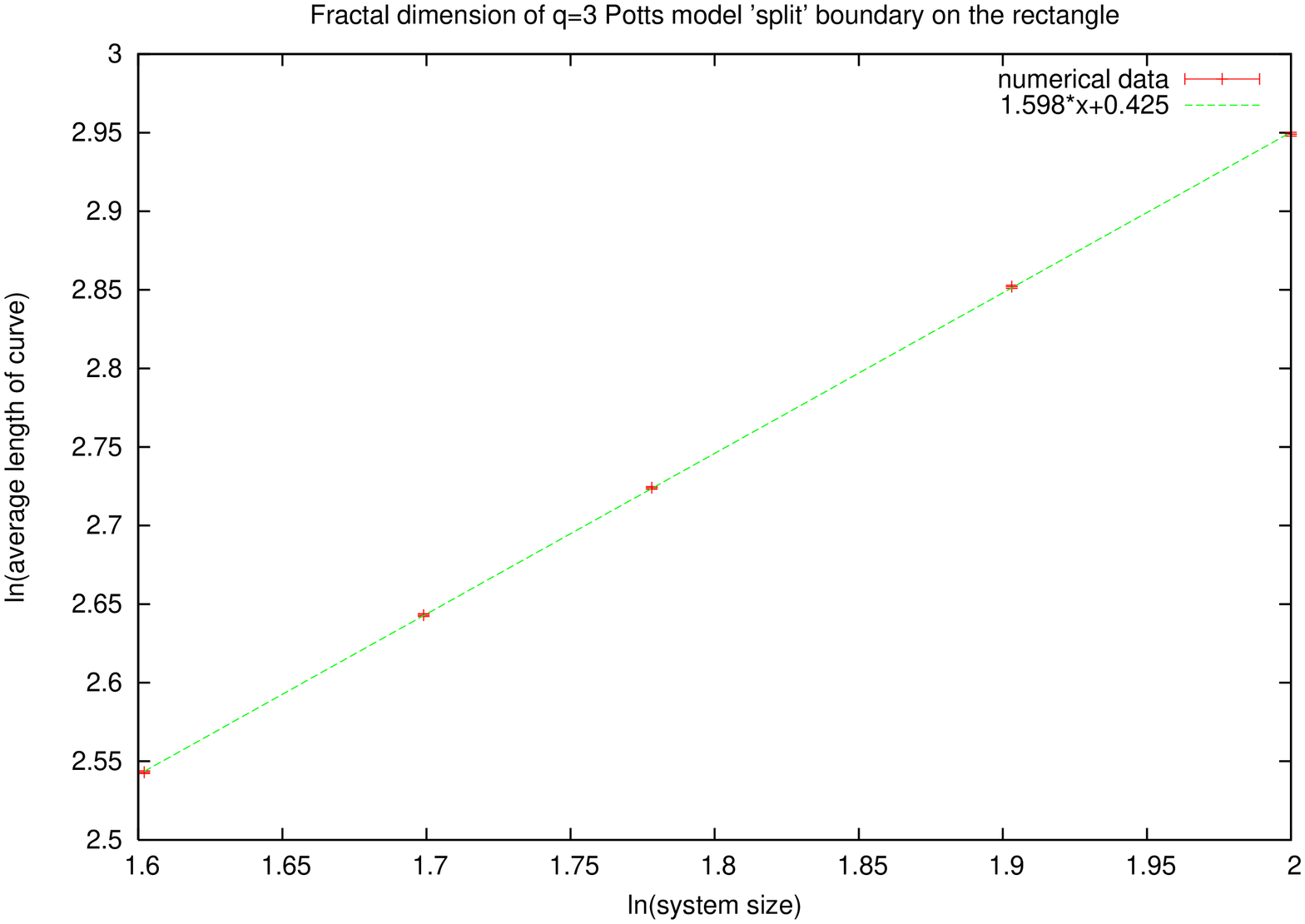, angle=270}}
    \caption{Fractal dimension of `composite' and `split' boundary
types for $q=3$ Potts model on the rectangle with fixed boundary
conditions. The system size is the length of the smaller side of the rectangle
in units of lattice spacing.}
    \label{Fig3qRectComplexSimpleFract}
  \end{center}
\end{figure}

An alternative set of boundary conditions are the `fluctuating'
boundary conditions described in section~\ref{SecPotts}. These
lead to the existence of two cluster boundary curves; the boundary
of the cluster containing one set of boundary spins and the
boundary of the cluster containing the other boundary spins.
Wherever the two curves overlap, this is labelled a `composite'
part of the curve. Where they are distinct they are labelled
`split' curves. In order to ascertain the fractal dimensions of
the two types of curves, the expected total length of the two
types of curves as a function of system size is plotted, see
figure~\ref{Fig3qRectComplexSimpleFract}.
$65,000$ independent samples are generated for each sample size.
The fractal dimension of the `composite' boundary type is
$d_{f}=1.022\pm 0.003$, which would correspond to SLE with
$\kappa=0.18\pm 0.02$. The reduced $\chi^2$ of this linear fit is
$9.7$ however, which is indicative of non-linearity in the
results. Closer investigation reveals convexity resulting from the
fractal dimension diminishing as system size increases. The
`split' boundary type has a larger fractal dimension of
$d_{f}=1.598\pm 0.008$, which would result from SLE with
$\kappa=4.78\pm 0.06$. This supports the hypothesis that the
`split' boundary types dominate in the scaling limit, resulting in
a pair of curves running between the boundary points. The value of
$\kappa$ is a little surprising; it is close to the dual value for
$\kappa=10/3$, namely $\kappa'=16/\kappa=4.8$.

\begin{figure}[thbp]
  \begin{center}
  \scalebox{0.35}{\epsfig{figure=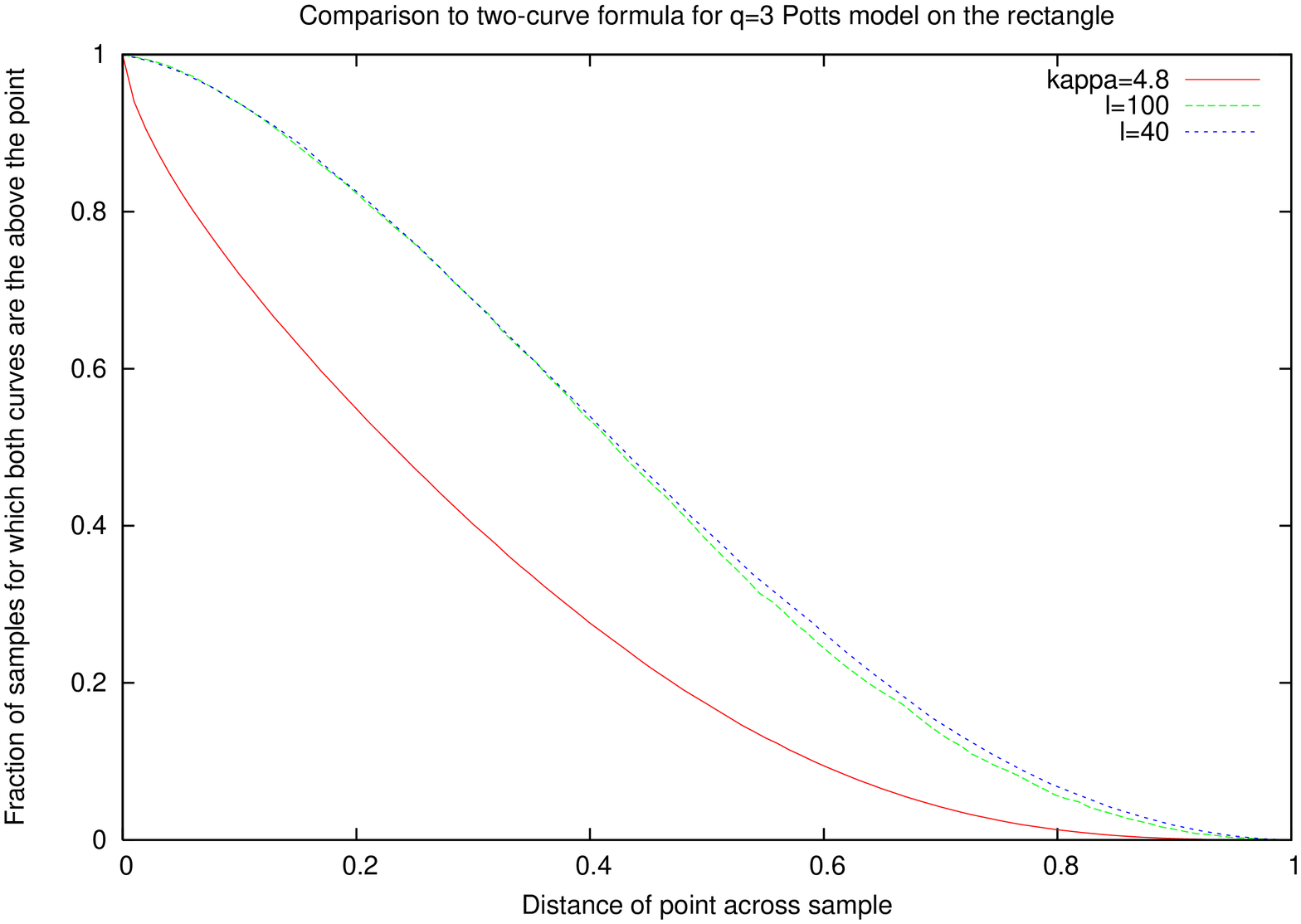, angle=270}}
  \scalebox{0.35}{\epsfig{figure=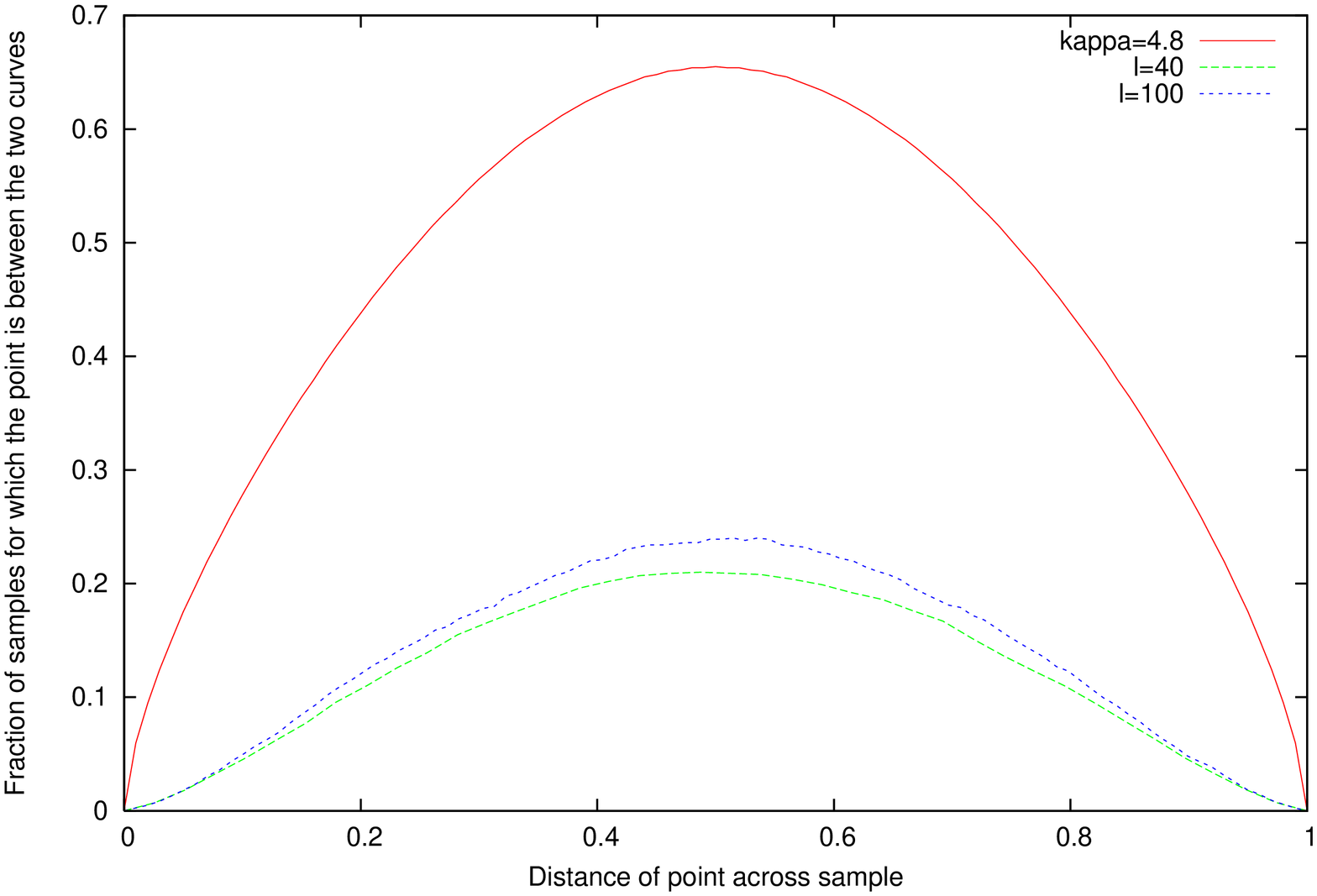, angle=270}}
    \caption{Comparison to the two-curve formula for the three-state Potts model on the
rectangle. The agreement with the prediction of
$\kappa=4.8$ from the fractal dimension is poor. Two
system sizes are shown, $l=40$ and $l=100$, where $l$ is the
smaller of the sides of the rectangle.}
    \label{Fig3qRect2CSchramm}
  \end{center}
\end{figure}

Figure~\ref{Fig3qRect2CSchramm} shows the agreement with the
generalisation of Schramm's formula to two curves, again
from $65,000$ independent samples. The path used is the same as
for the single curve case, namely the dashed line in
figure~\ref{FigIsRectGeom}. As a function of the distance of a
point upwards from the bottom boundary, we may determine the
expected fraction of the samples for which both cluster boundary
curves are above the point (top figure) and the fraction for which
the point is between the two curves (bottom figure). The fit to
the two-curve formula is poor for the lattice sizes
considered, as may have expected; for finite lattice size, there
is a non-zero contribution from `composite' cluster boundaries,
which reduce the proportion of points which lie between the two
curves and increase the proportion of points to one side of both
curves as a result.

\subsection{Ising model on the disc}\label{SubSecIsingdisc}
\begin{figure}[thbp]
  \begin{center}
  \scalebox{0.6}{\centerline{
    \epsfig{figure=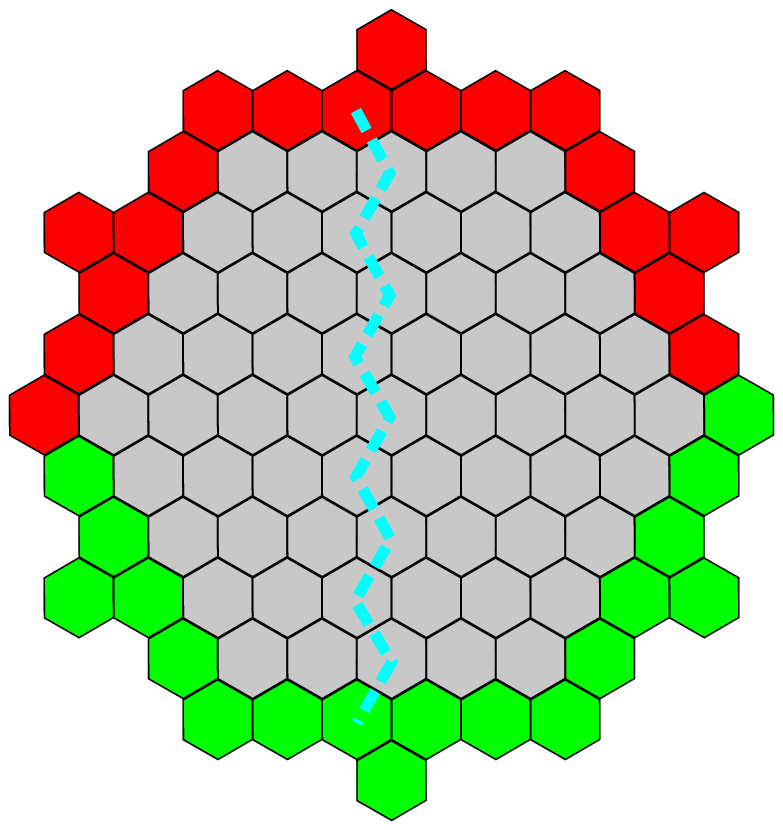}}}
    \caption{The geometry for the Ising model on the disc.
The figure would be classified as size $13$ as this is the number
of spins across the largest vertical diameter. Boundary spins are
fixed as shown and bulk spins left grey. The path used for the
comparison to Schramm's formula is shown as the dashed
cyan line.}
    \label{FigIsdiscGeom}
  \end{center}
\end{figure}

\begin{figure}[thbp]
  \begin{center}
  \scalebox{0.3}{\centerline{
    \epsfig{figure=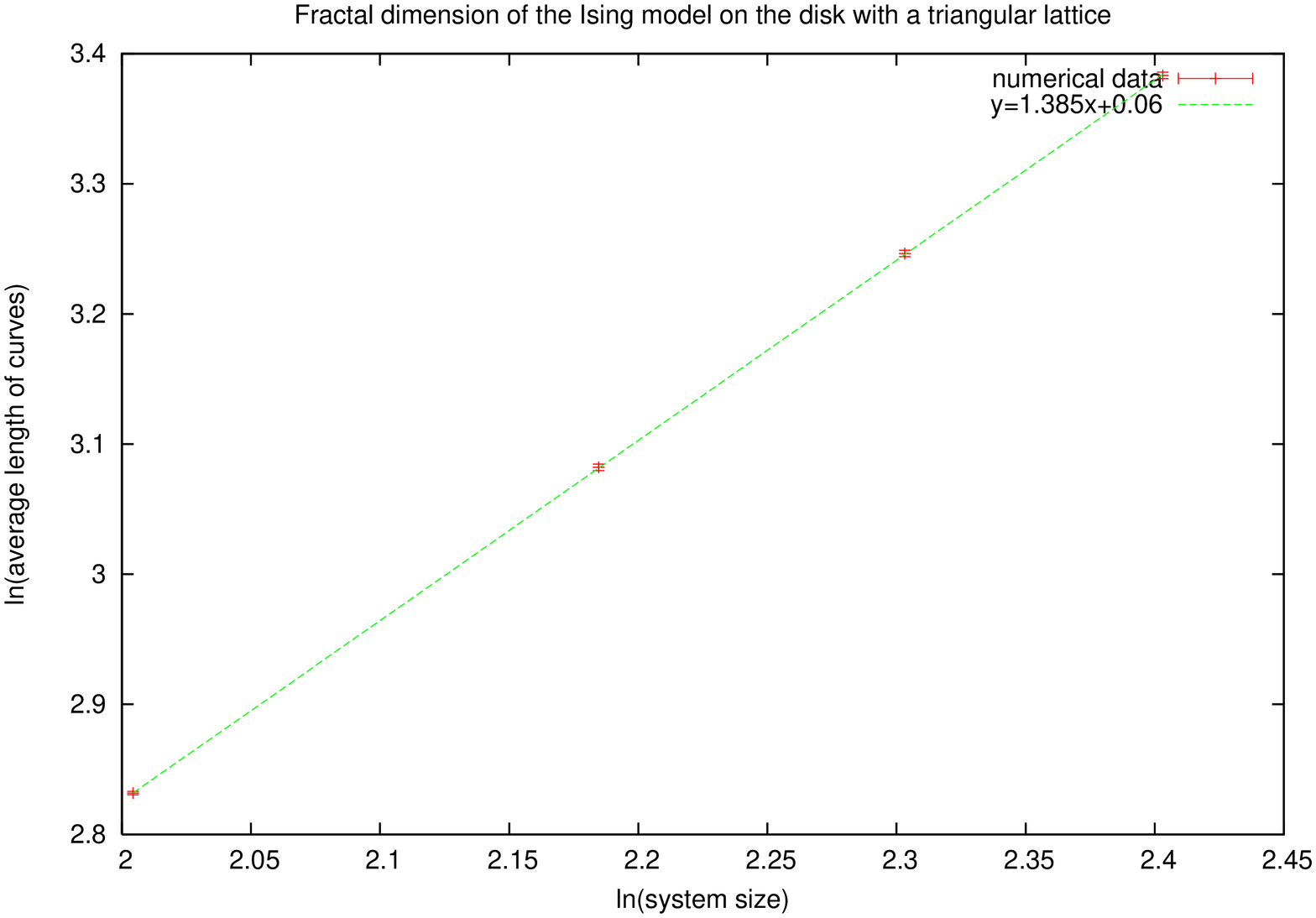, angle=270}}}
    \caption{The fractal dimension of the Ising model on the disc with a triangular lattice
from $10,000$ samples of each system size. The size of the system is the number of spins across
the central vertical diameter.}
    \label{Fig2qdiscFract}
  \end{center}
\end{figure}

The spins lie on the triangular lattice so that each spin has six
nearest neighbours and the spin boundaries lie on the edges of the
honeycomb lattice (the sides of the hexagons). Initially, four
system sizes are considered, characterised by the maximum number
of rows across the central vertical diameter: $101$, $153$, $201$
and $253$. Spins are included if the centre of the hexagon is a distance away from 
the centre of less
than or equal to the radius of this vertical column. The geometry
is shown in figure~\ref{FigIsdiscGeom} with the boundary
conditions used for the Ising model.
The fractal dimension of the curve in this geometry is
$d_{f}=1.38\pm 0.002$, which would correspond to SLE with
$\kappa=3.08\pm 0.02$, see figure~\ref{Fig2qdiscFract}. This
result is obtained from $10,000$ independent samples for each
system size.

\begin{figure}[thbp]
  \begin{center}
  \scalebox{0.35}{\epsfig{figure=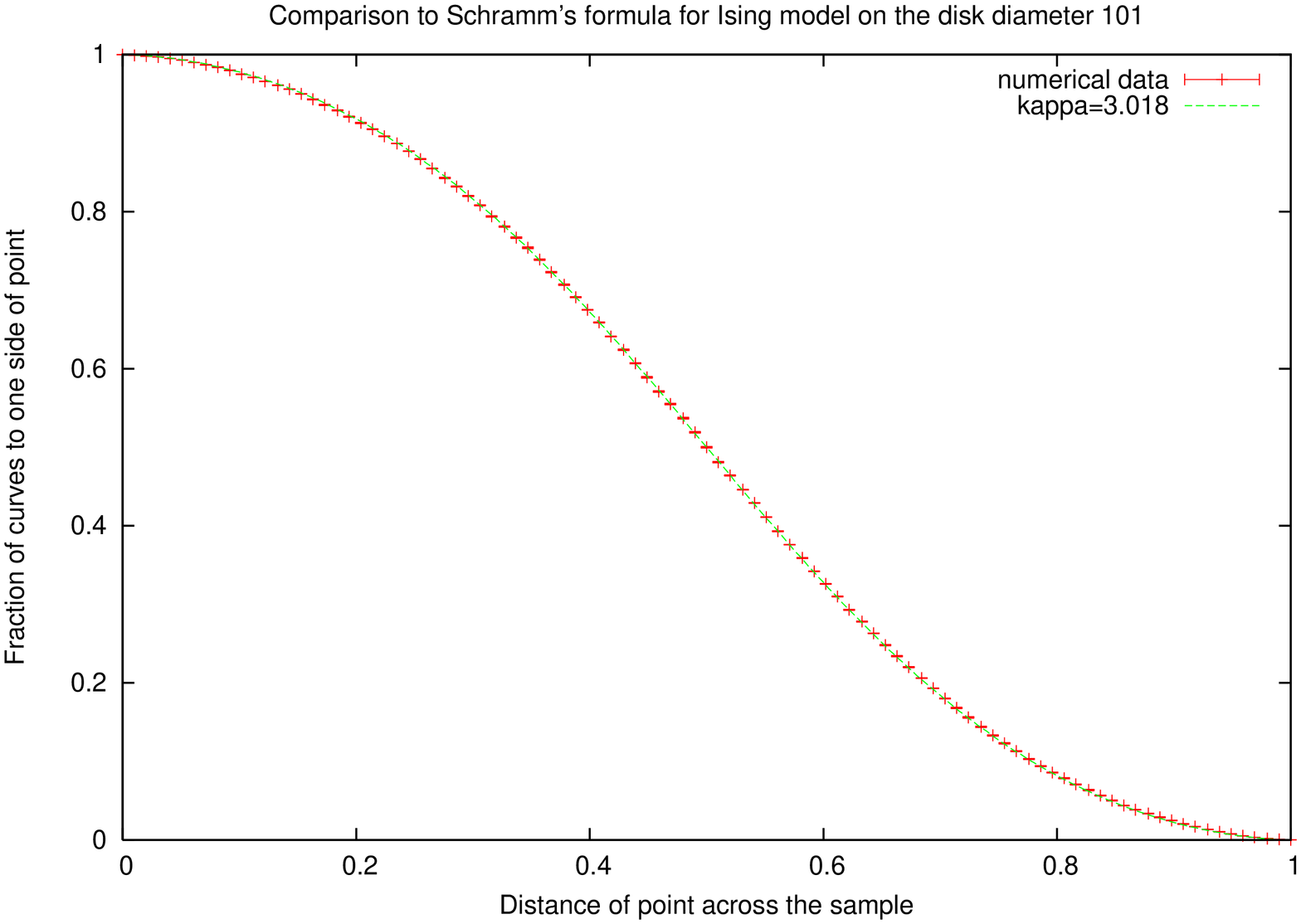, angle=270}}
  \scalebox{0.35}{\epsfig{figure=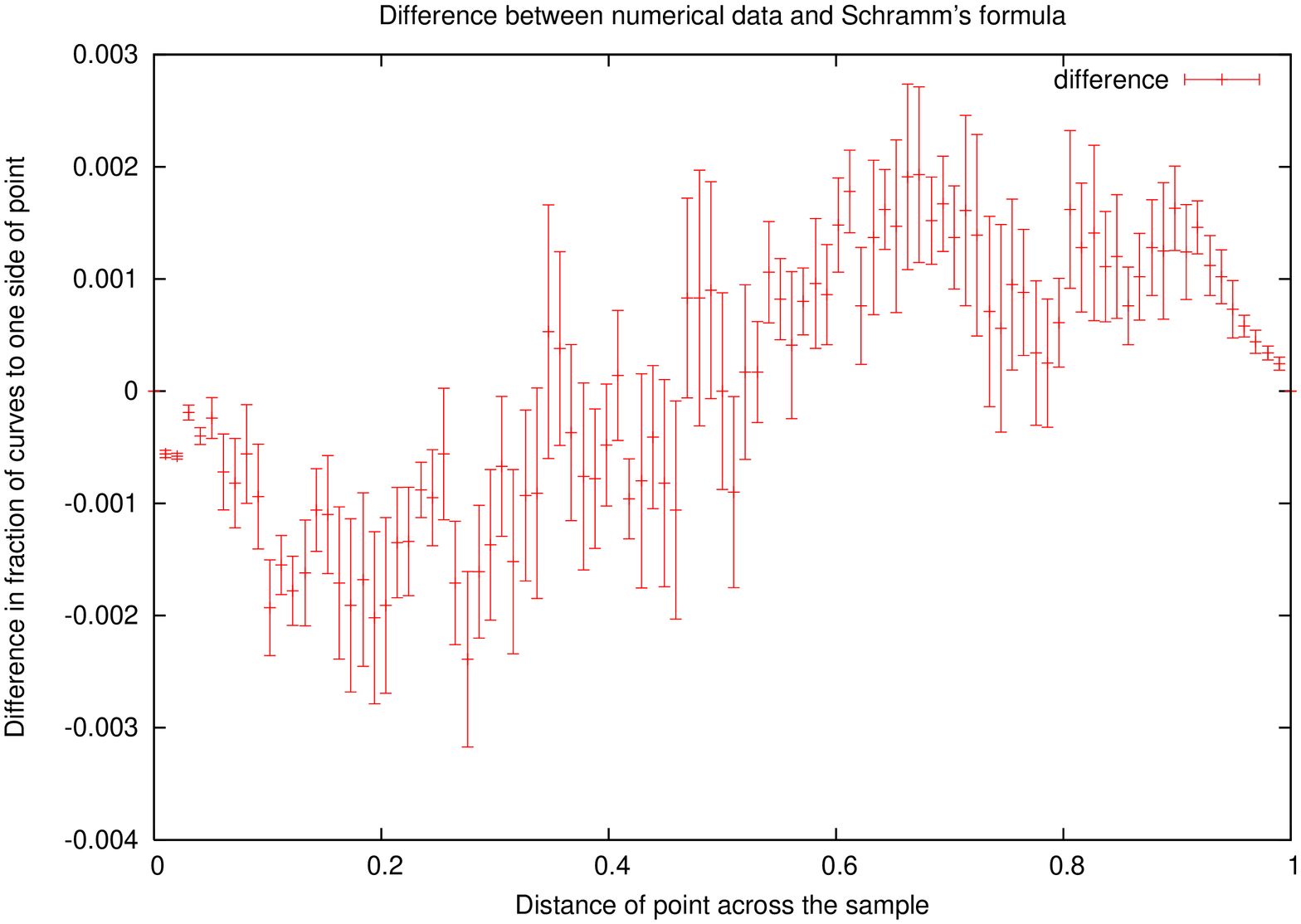, angle=270}}
    \caption{The best fit to Schramm's formula for the Ising model on the disc
with $101$ spins across the vertical diameter.. The agreement is to better than 1\%, see the lower figure.}
    \label{Fig2qdiscSchramm}
  \end{center}
\end{figure}

The fit to Schramm's formula is via the Mobius transformation from the upper half
plane to the disc
\begin{equation}\label{EqUHPtodisc}
  r(z)=-\frac{z-i}{z+i}\,.
\end{equation}
The value of $t$ in equation~\ref{EqSchramm} is related to the fractional distance
$0\leq s\leq 1$ upwards along the vertical diameter by the formula
\begin{equation}\label{EqTdisc}
  t=\frac{2s-1}{2s(1-s)}\,.
\end{equation}
The best fit for the largest system size considered would
correspond to SLE with $\kappa=3.018\pm 0.007$, as shown in
figure~\ref{Fig2qdiscSchramm}. The reason that only the largest
system size is shown is that the plot is quantitatively similar
for all system sizes. Over the range of system sizes tested, the
value of $\kappa$ fluctuates around a constant value.
This appears inconsistent with the hypothesis that the
value should be converging to $\kappa=3$ as system size increases,
but is likely to be a result of the small range of system sizes
considered. Larger systems are computationally demanding to
simulate, but would offer further insight into this problem.

\begin{figure}[thbp]
  \begin{center}
  \scalebox{0.35}{\centerline{
    \epsfig{figure=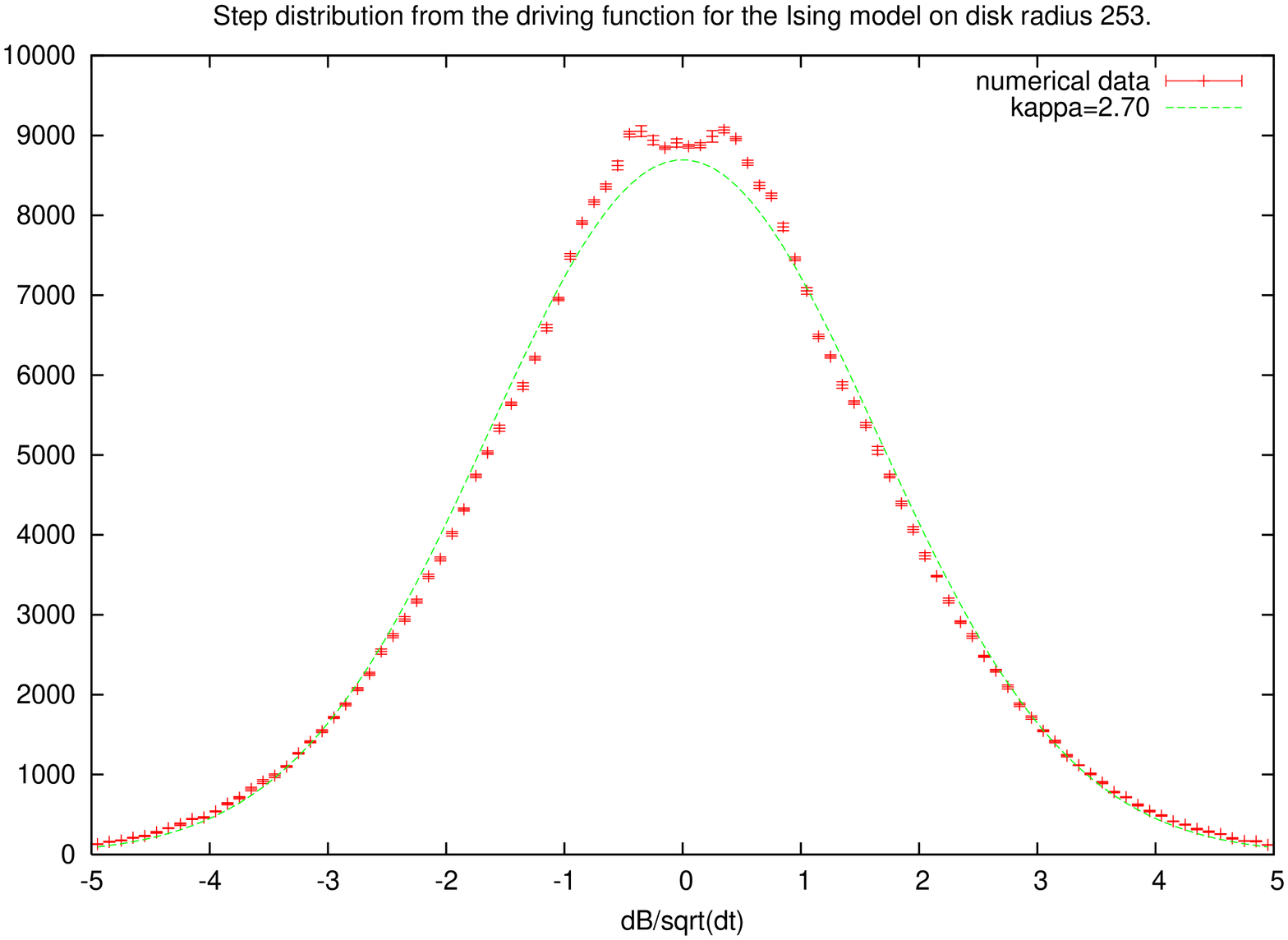, angle=270}}}
    \caption{The frequency of step distribution for the Ising model on the disc, diameter 253.
The plot takes data from $50,000$ independent samples. The y-scale
is an arbitrary measure of frequency.}
    \label{Fig2qdiscStep}
  \end{center}
\end{figure}

\begin{figure}[thbp]
  \begin{center}
  \scalebox{0.35}{\epsfig{figure=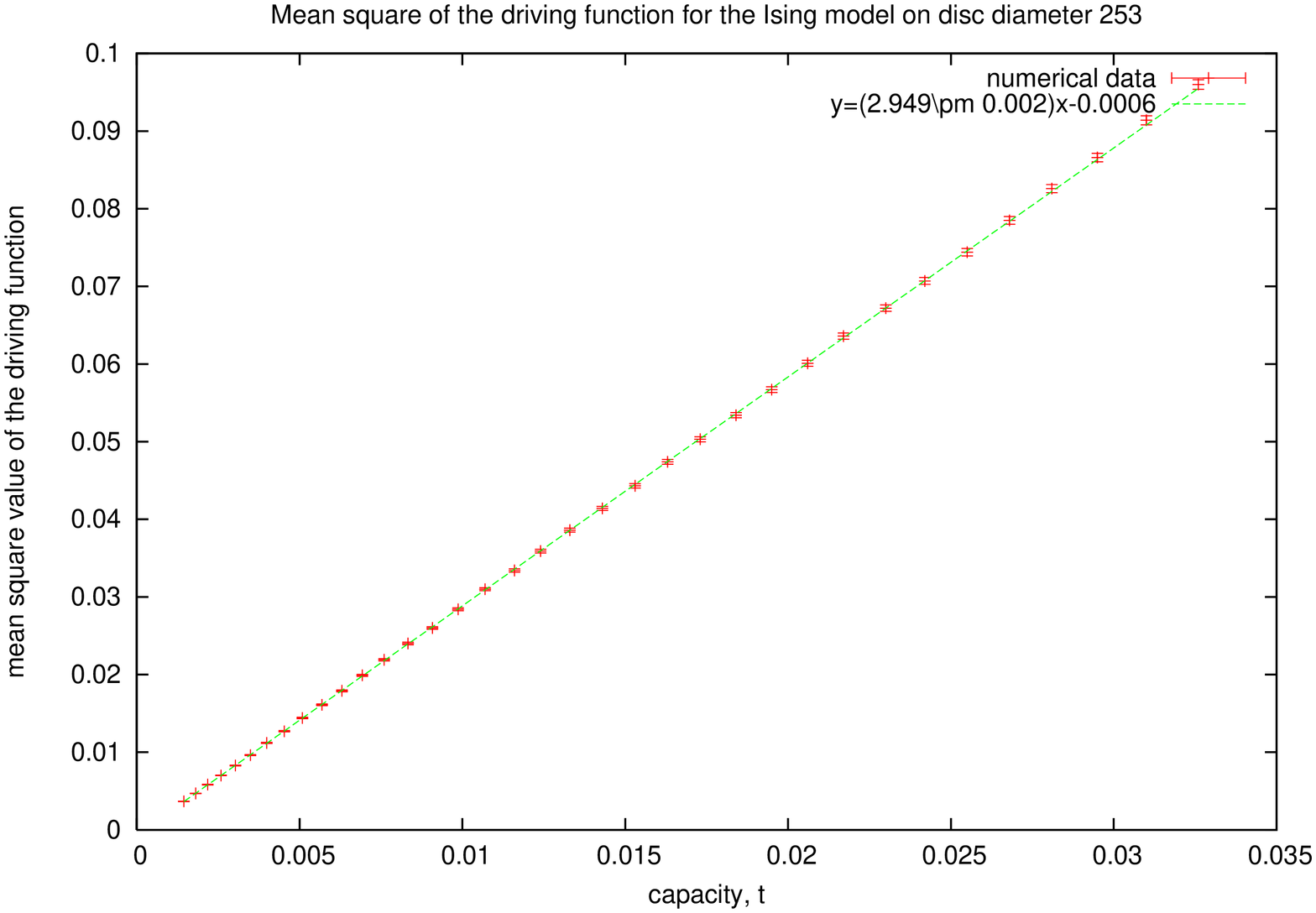, angle=270}}
  \scalebox{0.35}{\epsfig{figure=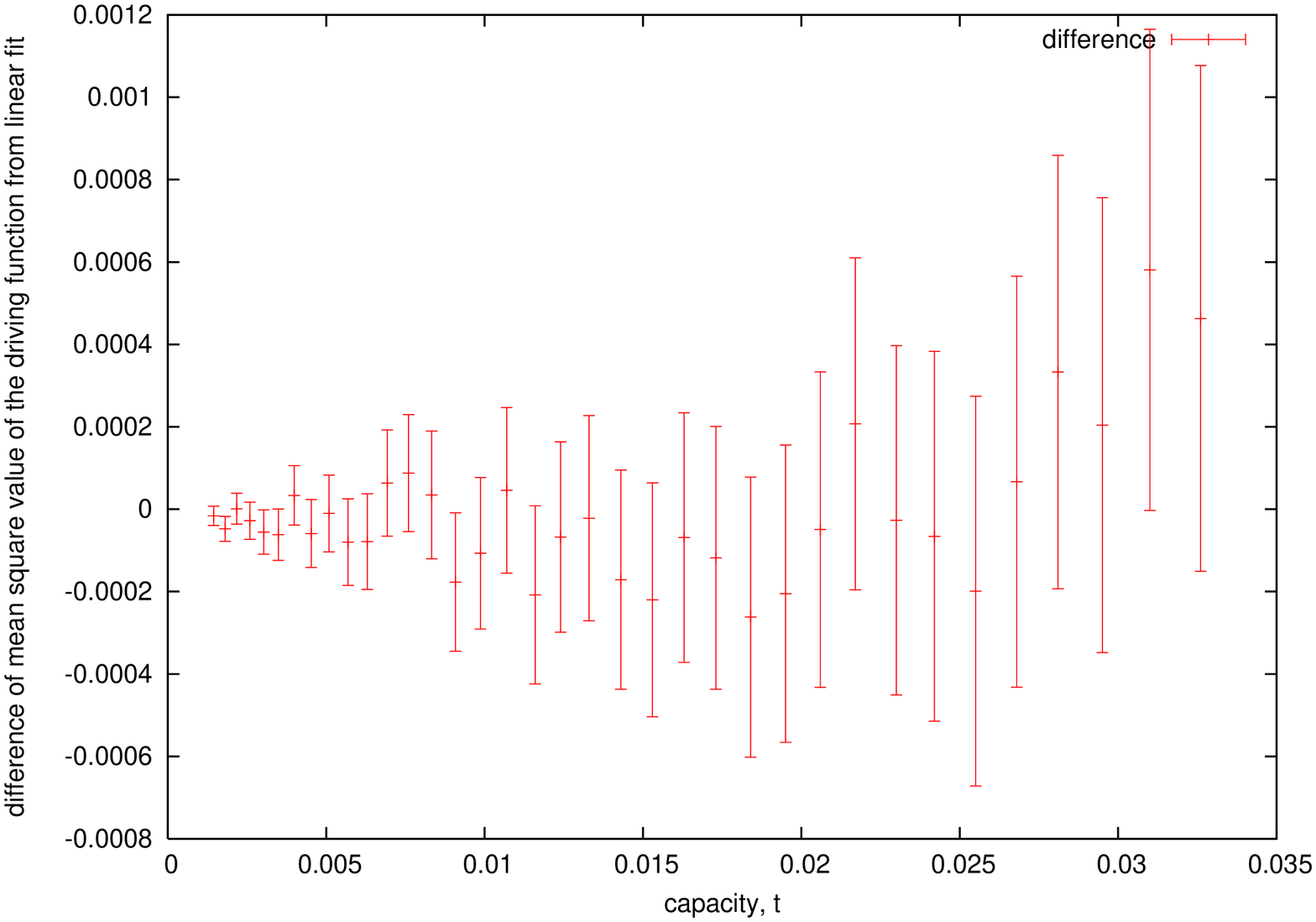, angle=270}}
    \caption{The expected value of $a_{t}^2$ plotted against capacity, $t$,
for the Ising model on the disc diameter $253$. The fit is
excellent, with reduced $\chi^2=0.01$. The data comes from
$40,000$ independent samples.}
    \label{Fig2qdiscLine}
  \end{center}
\end{figure}

\begin{figure}[thbp]
  \begin{center}
  \scalebox{0.35}{\centerline{
    \epsfig{figure=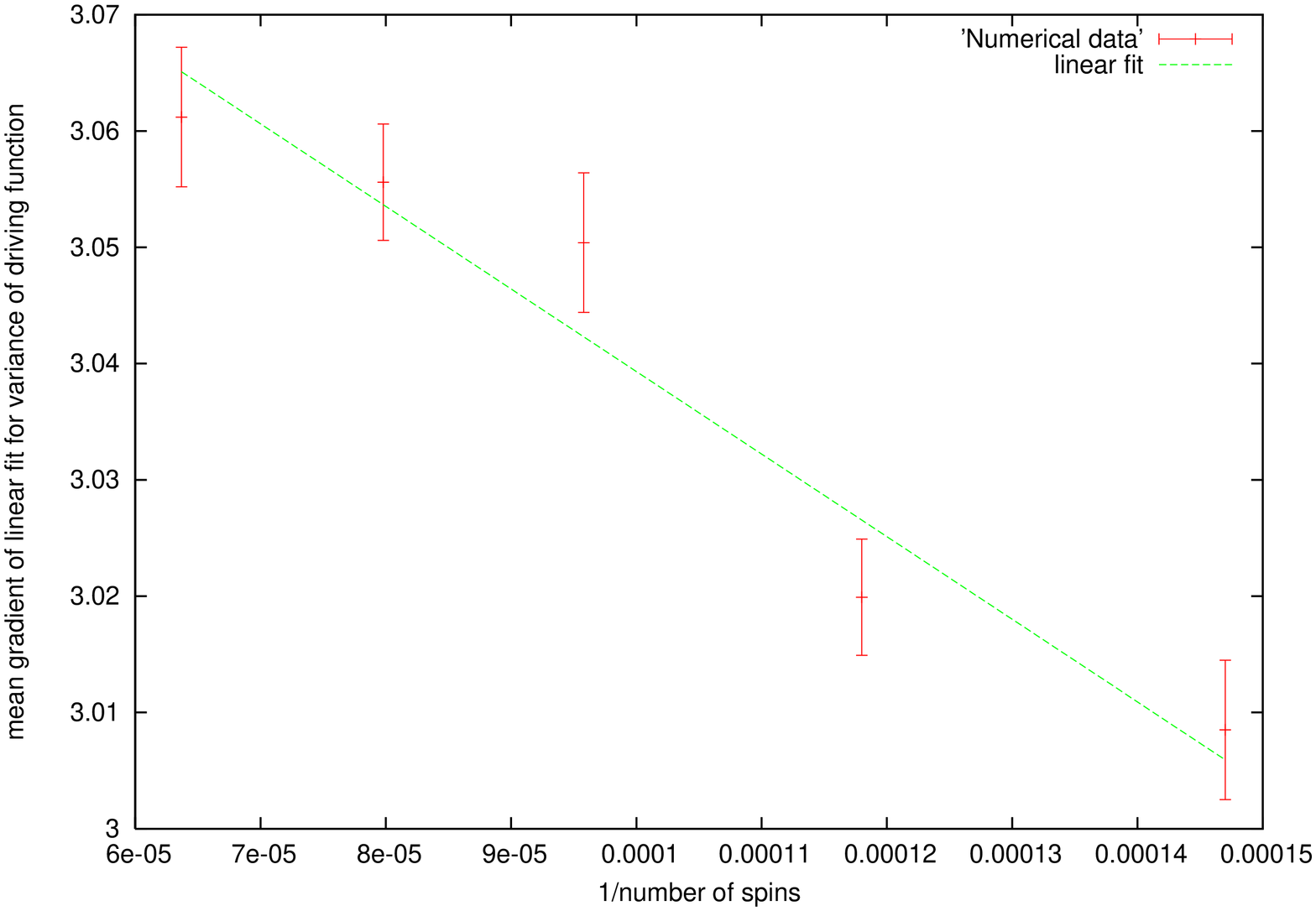, angle=270}}}
    \caption{The mean gradient of the linear fit of $a_{t}^2$ to $t$
for the Ising model as a function of system size. The fit shown is
linear, which is clearly inadequate; the approach to the scaling
limit is not a simple power law at these scales.}
    \label{FigIsdiscTend}
  \end{center}
\end{figure}

The third test against SLE is to extract driving functions
for each curve and to examine their statistics as a function of
system size. This process is described in
section~\ref{SecDrivingFunction}. An example, for diameter $253$,
is shown in figure~\ref{Fig2qdiscStep}. The results are from
$50,000$ independent samples. For each sample, only the part of the curve
corresponding to capacity $t\leq 1/3$ was used, since the error
introduced by the series of conformal maps may be expected to grow
exponentially. This would result in inaccurate values for the
driving function from later points on the curve.
The form should be Gaussian, but the finite lattice is apparently
responsible for a bi-modality for small steps $\delta
B_{t}/\sqrt{\delta t}$. A second test is to examine the expected
squared value of the driving function as a function of capacity,
$t$. Each system size has a linear fit with a gradient which would
correspond to $\kappa$. The results for diameter $253$ is shown in
figure~\ref{Fig2qdiscLine}. The linear fit is excellent, with
reduced $\chi^2=0.01$.
The value of this best fit as a function of the inverse system
size is plotted in figure~\ref{FigIsdiscTend}. The approach to the
scaling limit appears to be complicated; it is not a simple power
law at these lattice sizes.

Smaller system sizes and larger numbers of independent samples are used
for this test, as large numbers of samples are required to obtain
results to a desirable accuracy. The table below shows the system
sizes used, with the corresponding values of
the inverse number of spins and number of samples generated.

\begin{tabular}{|c|c|c|}
\hline  System size & Inverse number of spins & Number of samples
used \\\hline
  101 & $1.47\times 10^{-4}$ & 760,000 \\
  113 & $1.18\times 10^{-4}$ & 920,000 \\
  125 & $9.58\times 10^{-5}$ & 720,000 \\
  137 & $7.98\times 10^{-5}$ & 1,040,000 \\
  153 & $6.37\times 10^{-5}$ & 580,000 \\ \hline
\end{tabular}
\subsection{$q=3$ Potts model on the disc}\label{SubSec3qdisc}
\begin{figure}[thbp]
  \begin{center}
  \scalebox{0.35}{\centerline{
    \epsfig{figure=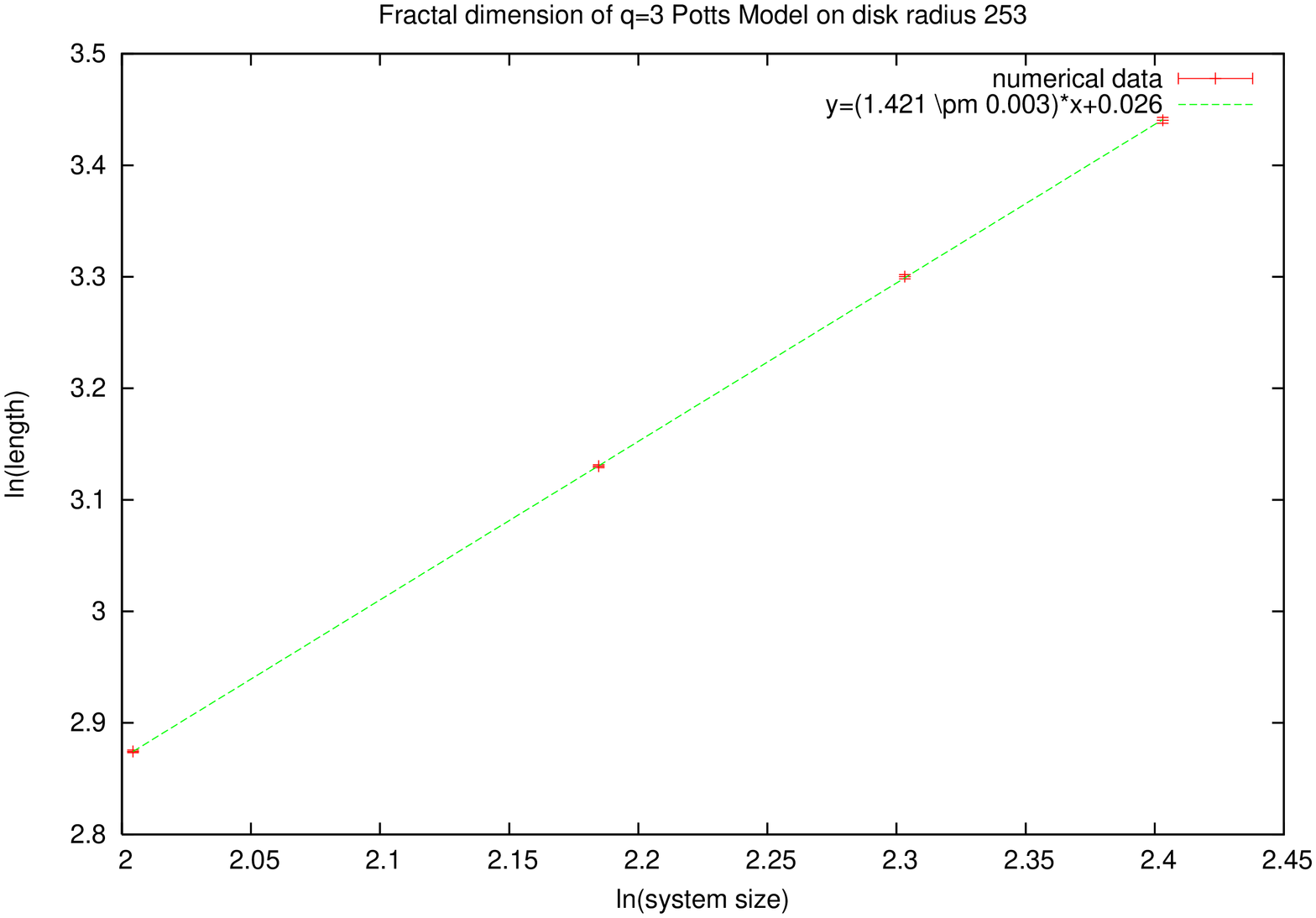, angle=270}}}
    \caption{The fractal dimension of the $q=3$ Potts model on the disc.
System size is the number of spins across the central vertical diameter.}
    \label{Fig3qdiscFract}
  \end{center}
\end{figure}

\begin{figure}[thbp]
  \begin{center}
  \scalebox{0.35}{\epsfig{figure=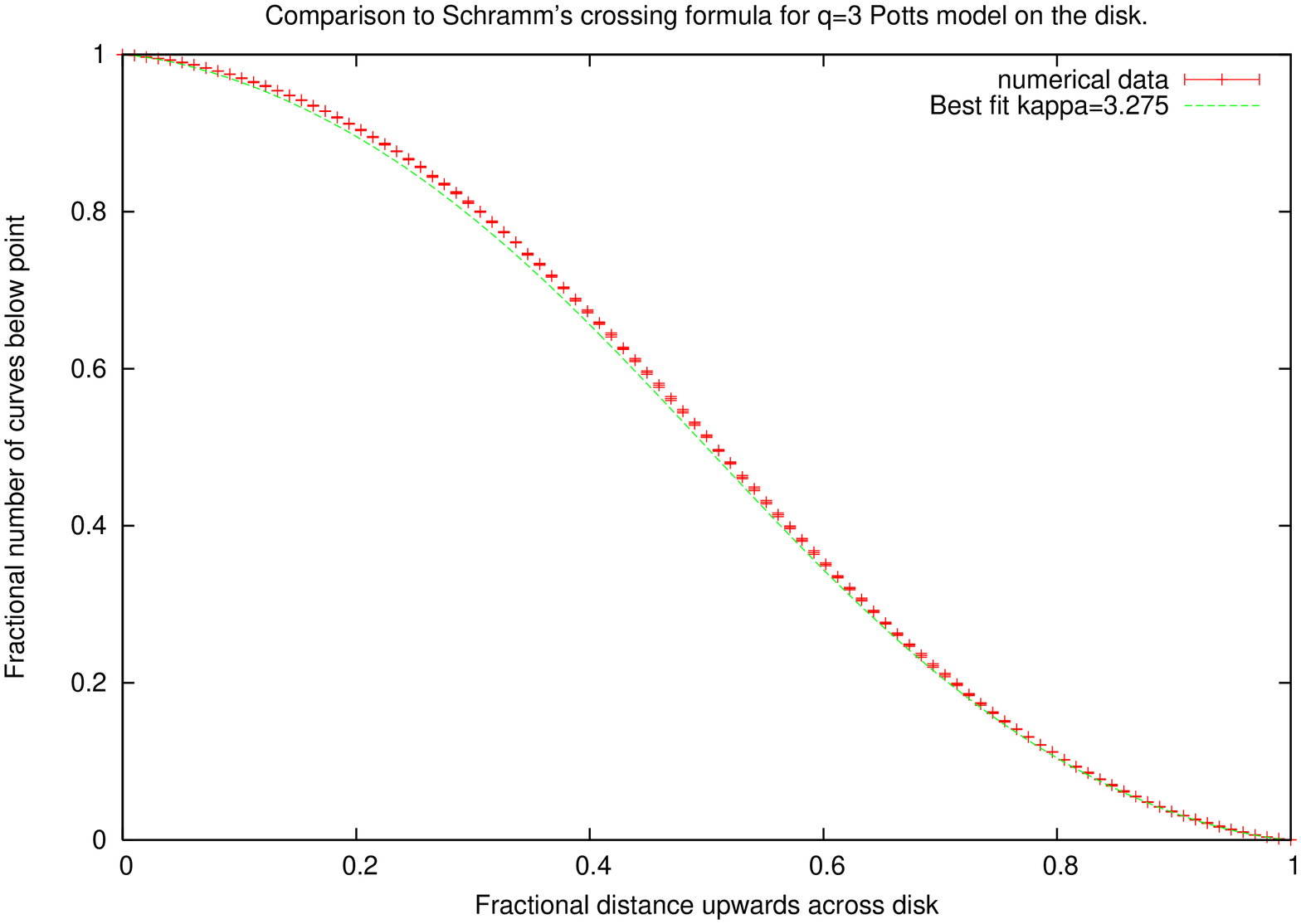, angle=270}}
  \scalebox{0.35}{\epsfig{figure=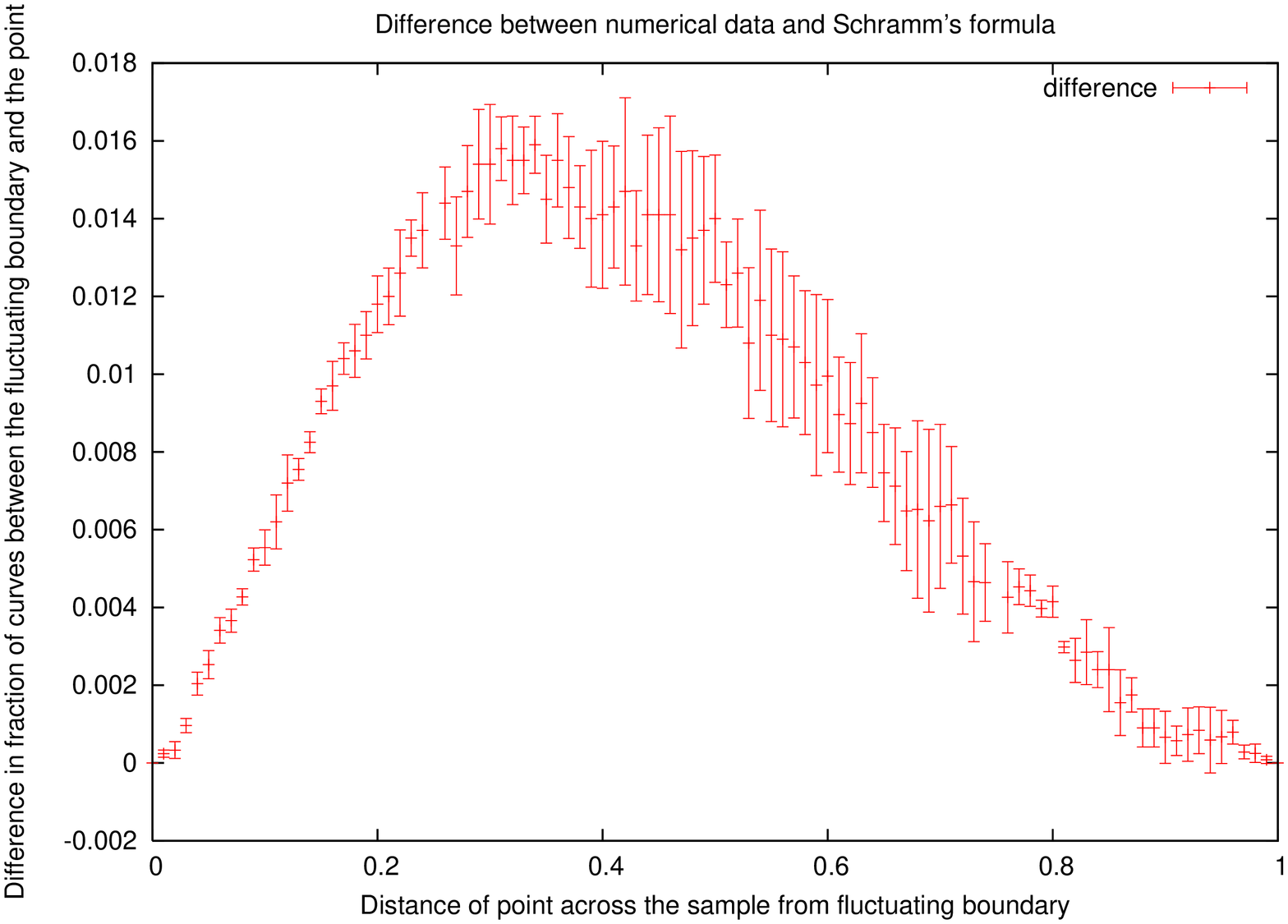, angle=270}}
    \caption{Best fit to Schramm's formula for the $q=3$ Potts model on the disc diameter $101$.
The same asymmetry is seen as for the model on the rectangle.}
    \label{Fig3qdiscSchramm}
  \end{center}
\end{figure}

\begin{figure}[thbp]
  \begin{center}
  \scalebox{0.35}{\epsfig{figure=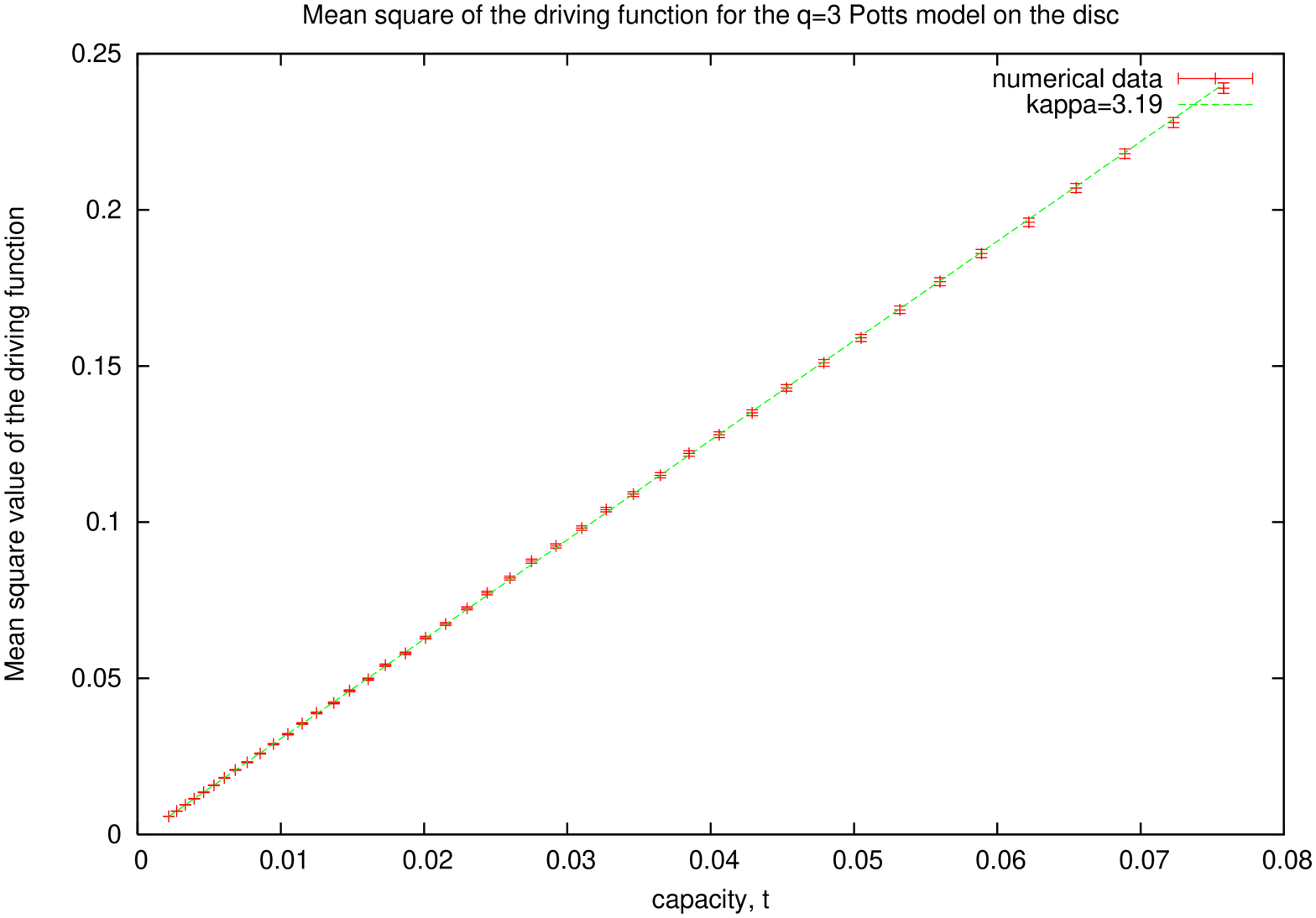, angle=270}}
  \scalebox{0.35}{\epsfig{figure=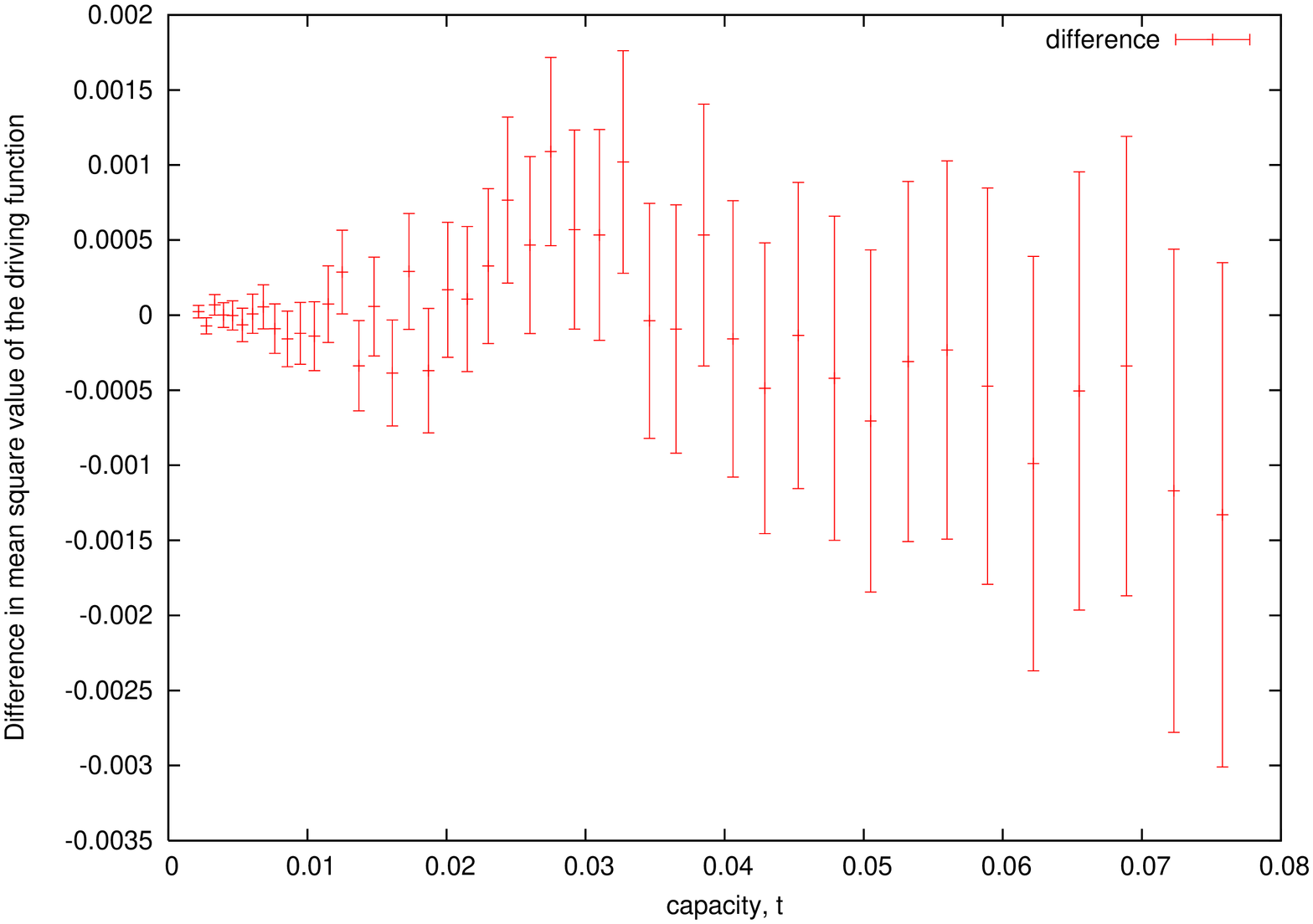, angle=270}}
    \caption{The mean square value of the driving function as a function of capacity, $t$.
The system is the disc of diameter $201$ and $40,000$ independent
samples are used. Agreement with the linear fit is excellent.}
    \label{Fig3qdiscFit}
  \end{center}
\end{figure}

\begin{figure}[thbp]
  \begin{center}
  \scalebox{0.35}{\centerline{
    \epsfig{figure=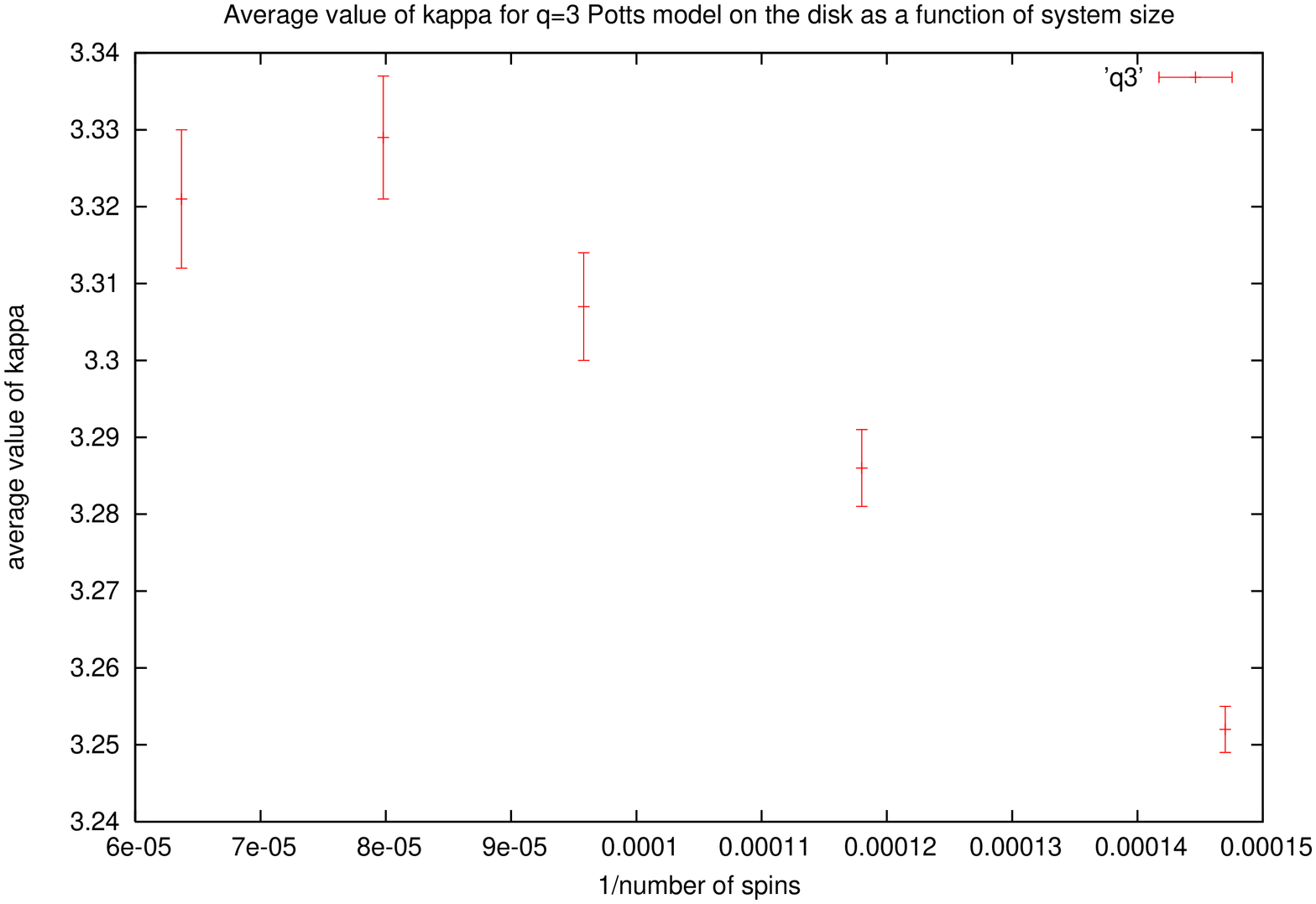, angle=270}}}
    \caption{The gradient of the linear fit to the mean square value of the driving
function against capacity, plotted against inverse system size.
The approach to the scaling limit is not a simple power law.}
    \label{Fig3qdiscTend}
  \end{center}
\end{figure}

Firstly, we examine the case of `fluctuating' boundary conditions,
such that a single cluster boundary propagates through each
sample. System sizes are again characterised by the number of
spins along the central vertical diameter. System sizes considered
are $101$, $153$, $201$ and $253$. The fractal dimension, obtained
as usual by determining the expected curve length as a function of
system size, is plotted in figure~\ref{Fig3qdiscFract}.
The value for the fractal dimension obtained is $d_{f}=1.4216\pm
0.0007$, which would correspond to SLE with $\kappa=3.373\pm
0.006$. This comes from considering $10,000$ independent samples
of each system size. The fit to Schramm's formula, as for
the Ising model, fluctuates with system size across the range of
sample sizes considered. For the largest system size considered,
the best fit is to the predictions from SLE with $\kappa=3.275\pm
0.003$. This is shown in figure~\ref{Fig3qdiscSchramm}.
The driving function may be obtained by considering the series of
conformal maps which undo the curve up to capacity, $t$, as for
the Ising model in the previous subsection. A similar agreement
with a Gaussian fit is again seen, with the same characteristic
bi-modal behaviour at small $\delta B_{t}/\sqrt{\delta t}$. It is
more instructive to examine $\la B_{t}^{2}\ra$ as a function of
$t$ (the variance of the distribution) and to extract the gradient
of the linear fit. Each system size shows linear behaviour to a
high degree of accuracy. A typical plot is as in
figure~\ref{Fig3qdiscFit}, which has reduced $\chi^2=0.01$.
The values of the gradient of the linear fit to $\la a_{t}^2\ra$
against capacity, as a function of inverse system size, is shown
in figure~\ref{Fig3qdiscTend}. As for the Ising model, smaller
system sizes are used for this test because large number of
independent samples are required to keep the error manageable. As
for the Ising model, the system does not show a power law approach
to the scaling limit at these system sizes.

\begin{tabular}{|c|c|c|}
\hline  System size & Inverse number of spins & Number of samples
used \\\hline
  101 & $1.47\times 10^{-4}$ & 800,000 \\
  113 & $1.18\times 10^{-4}$ & 800,000 \\
  125 & $9.58\times 10^{-5}$ & 800,000 \\
  137 & $7.98\times 10^{-5}$ & 640,000 \\
  153 & $6.37\times 10^{-5}$ & 720,000 \\ \hline
\end{tabular}

\begin{figure}[thbp]
  \begin{center}
  \scalebox{0.35}{\centerline{
    \epsfig{figure=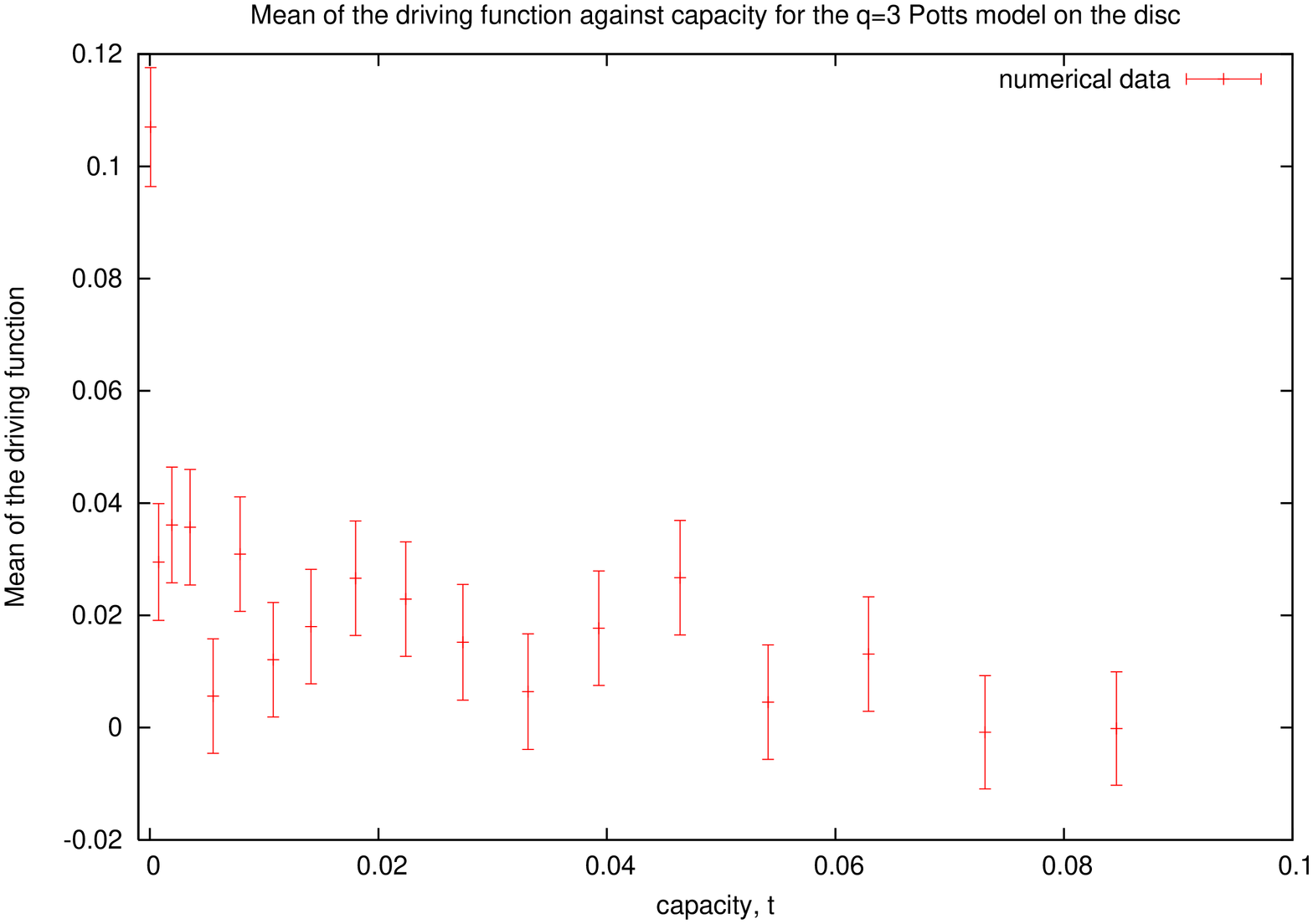, angle=270}}}
    \caption{Mean of the driving function as a function of capacity for the $q=3$
Potts model on the disc diameter 253.}
    \label{FitMeanDis}
  \end{center}
\end{figure}

Recall that, by comparison to the fit to Schramm's
formula, the spin boundary curve of the $q=3$ Potts model with
`fluctuating' boundary conditions displays asymmetry. The curve is
more likely to pass close to the fluctuating boundary than is
predicted by SLE. This asymmetry is not surprising; the boundary
conditions are not reflection symmetric, as they are in the Ising
model. However, this asymmetry must vanish in the scaling limit if
the curve is to become SLE. Figure~\ref{FitMeanDis} shows the mean
of the driving function for the curve on the disc of diameter
$253$ as a function of capacity, $t$. It is clear that the
asymmetry is a result of the behaviour of the curve at small $t$.
This is encouraging, since it suggests that this is a boundary
effect, whose contribution would therefore become irrelevant in
the scaling limit.

\begin{figure}[thbp]
  \begin{center}
  \scalebox{0.35}{\epsfig{figure=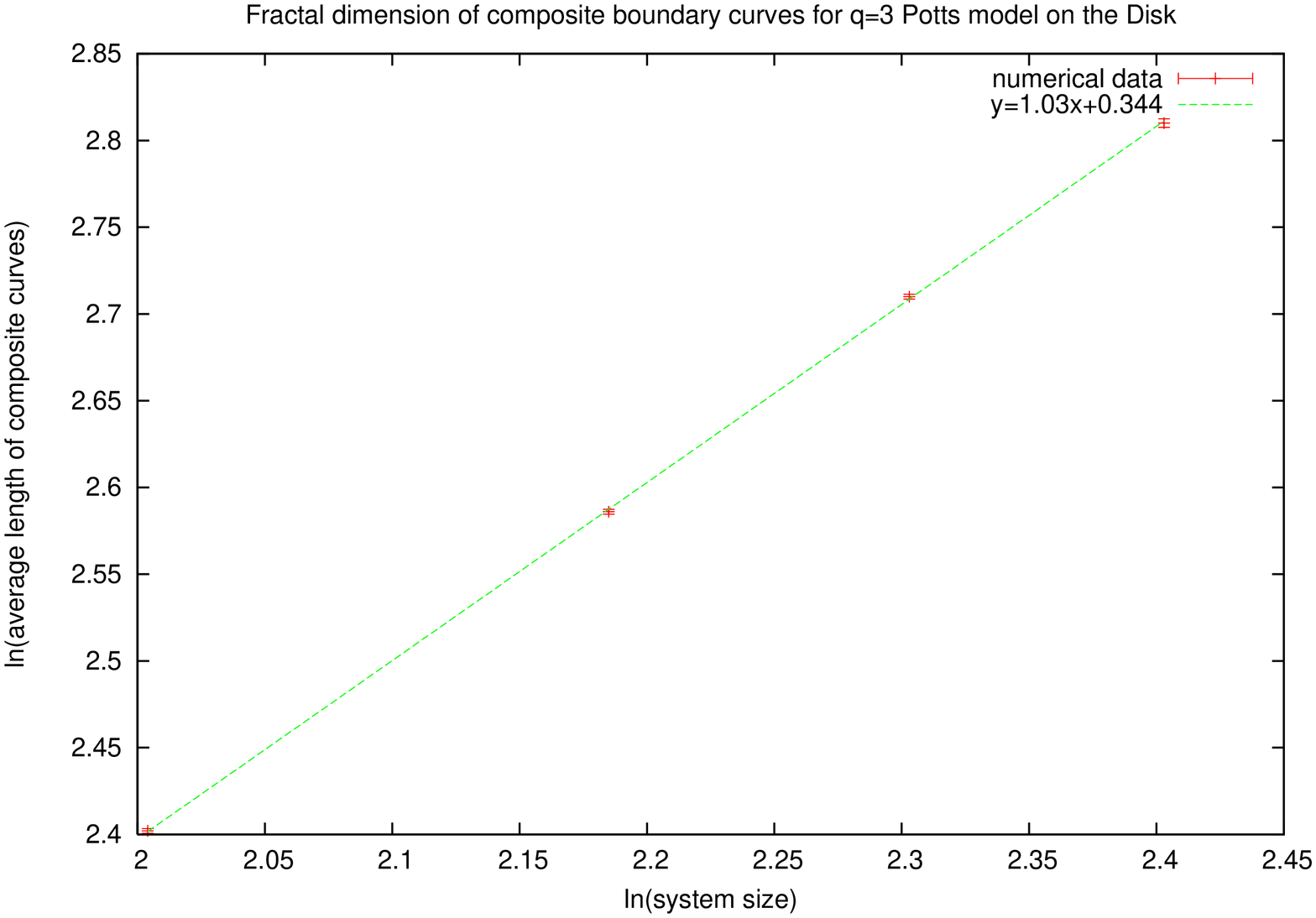, angle=270}}
  \scalebox{0.35}{\epsfig{figure=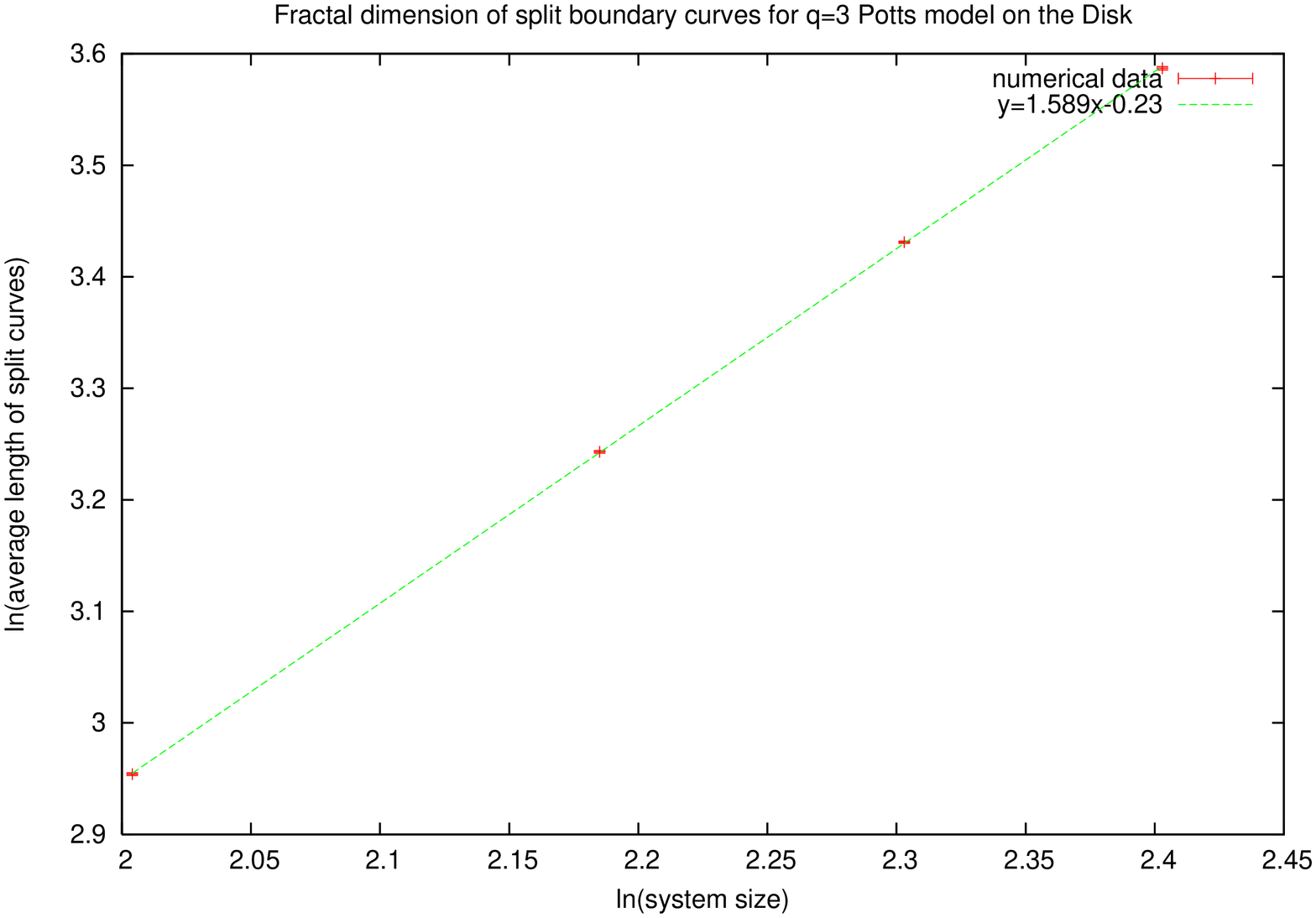, angle=270}}
    \caption{The fractal dimensions of the `composite' (top figure) and `split' (bottom figure)
boundaries for the $q=3$ Potts model on the disc.}
    \label{FigdiscFixed}
  \end{center}
\end{figure}

The case of `fixed' boundary conditions may be tested for fractal
dimension of the two types of boundary, the `composite' and
`split' boundaries. A plot of the fractal dimension of each of
these is shown in figure~\ref{FigdiscFixed}. $40,000$ independent
samples of each system size were used. The fractal dimension of
`composite' curves is not constant but diminishing as system size
increases, just as was seen for the model on the rectangle. The
fractal dimension of the `split' curves $d_{f}=1.589\pm 0.005$
would correspond to SLE with $\kappa=4.71\pm 0.04$. The fractal
dimension of the the `composite' curves is $d_{f}=1.03\pm 0.01$
($\kappa=0.24\pm 0.08$).
\newpage
\section{Summary and Conclusions}\label{SecSummary}
Recall our hypothesis, that the scaling limit of spin cluster boundaries in the
three-state Potts model is SLE with $\kappa=10/3$.
The numerical results from the simulations with `fluctuating' boundary conditions,
and hence with single curves propagating across the samples,
shows encouraging support for this hypothesis.
The data should be compared to that from the Ising
model, which has been proven to correspond to $SLE_{3}$. The
results are summarised in the table below for the tests of the
fractal dimension and best fit to Schramm's formula, both
on the rectangle and the disc.

\begin{tabular}{|c|c|c|}
\hline
  Rectangle & Ising & $q=3$ \\
\hline
  Fractal dimension & $2.976\pm 0.02$ & $3.192\pm 0.02$ \\
  Schramm's formula & $3.02\pm 0.02$ & $3.235\pm 0.01$ \\ \hline
\end{tabular}

\begin{tabular}{|c|c|c|}
\hline
  disc & Ising & $q=3$ \\
\hline
  Fractal dimension & $3.08\pm 0.02$ & $3.373\pm 0.006$ \\
  Schramm's formula & $3.018\pm 0.007$ & $3.275\pm 0.003$ \\
\hline
\end{tabular}

The value for the fractal dimension is the more robust result,
since it incorporates data from the whole range of system sizes. Notice
that the known value of $\kappa=3$ lies just outside the range obtained
from the fractal dimension of the numerical simulations on the rectangle.
The values obtained are a little too low for the Ising model, but also
for the three-state Potts model, compared to our prediction of $\kappa=10/3$.
However, on the disc, the Ising model shows a
fractal dimension which is larger than the known value. In this
case, the fractal dimension of the Potts model is also larger than
our prediction. Qualitative agreement with Schramm's
formula for a single system size ($100\times 300$ on the rectangle
and diameter $253$ on the disc) is also encouraging.
We have demonstrated that the asymmetry
of the curve (it is more likely to be close to the fluctuating
boundary than the prediction from SLE) is a result of the
behaviour of the curve at small capacity, as may be seen from a
plot of the mean of the driving function against capacity. This
asymmetry may be assumed, therefore, to be a boundary effect,
which would be irrelevant in the scaling limit. This infers that reflection
symmetry is recovered in the scaling limit, consistent with the curves being SLE.

Since the results from the variance of the driving functions do
not appear to approach the scaling limit as a simple power law, it is not clear how
to extract a value for $\kappa$ from the data. If we simply
average over the values obtained from all system sizes considered,
we find $\kappa=3.04$ for the Ising model and $\kappa=3.30$ for the $q=3$
Potts model. This figure is, of course, an arbitrary one and is
supposed only to give a guide to the order of magnitude of the
variance in the scaling limit. The fact that this figure is close
to the predicted value is, however, encouraging.

We also considered the three-state Potts model with fixed boundary
conditions, which leads to a pair of cluster boundaries,
coinciding in some places. The results for $\kappa$ from the
fractal dimension of these two curve types on the rectangle and
the disc are summarised below.

\begin{tabular}{|c|c|c|}
\hline
  2 Curves & Composite & Split \\
\hline
  Rectangle & $0.18\pm 0.02$ & $4.78\pm 0.06$ \\
  disc & $0.24\pm 0.08$ & $4.71\pm 0.04$ \\ \hline
\end{tabular}

The `split' boundaries have a larger fractal
dimension. This is consistent with the hypothesis that they
dominate in the scaling limit, supported by the irrelevance of the 4-leg
operator. For the sample sizes considered, the agreement with the
two-curve formula (the generalisation of Schramm's formula to two-curve SLE) is
poor. This is unsurprising, as the presence of `composite' cluster boundaries reduces
the frequency of points being between the two curves and
redistributes this to the probability that a point is to one side
of `both' curves.

Acknowledgements: The authors would like to thank Benjamin Doyon,
Tom Kennedy, Valentina Riva, Federico Camia and David Wilson for helpful discussions and encouragement.
This work was supported in part by EPSRC Grants
GR/R87712/01 and EP/D050952/1. It was carried out in part at the program on
`Stochastic Geometry' at the Kavli Institute for Theoretical Physics, whose hospitality is
gratefully acknowledged.
\newpage
\appendix
\section{The Fortuin-Kastelyn cluster representation}
There exists a well known mapping of the partition function to a
cluster model, which will be referred to in the appendix
describing the Wolff algorithm.
Up to a multiplicative factor, the partition function is
\begin{equation}\label{FK}
Z=\tr{Tr}\prod_{r,r'}(1-p+p\delta_{s(r),s(r')})\,,
\end{equation}
with $e^{-\beta J}=1-p$. The product may be expanded into a sum of
$2^{N}$ terms ($N$ is the number of nearest neighbours) by
choosing either $(1-p)$ or $p\delta_{s(r),s(r')}$ from each
bracket. If a line is drawn on the edge between each pair of spins
for which the term $p\delta_{s(r),s(r')}$ was chosen, the
partition function is seen to correspond to a sum over all graphs
on the edges of the original lattice. The spins in each connected
cluster are constrained to take the same value, so taking the
trace leads to
\begin{equation}
Z=\sum_{G}p^{|G|}(1-p)^{|\overline{G}|}Q^{||G||}\,.
\end{equation}
The sum is over all graphs, $|G|$ is the number of edges in a
given graph, $|\overline{G}|$ is the number in the complement and
$||G||$ is the number of connected components, known as
Fortuin-Kastelyn (FK) clusters.

\section{The Wolff algorithm}\label{SecWolffAlgorithm}
The Monte Carlo algorithm used throughout to generate samples was
the Wolff algorithm~\cite{Wolff,SwWang}. It is a Monte Carlo
algorithm with the following update procedure:
\begin{enumerate}
\item Pick a lattice site at random.
\item Visit all neighbouring sites, adding them to the cluster with a probability $p$ if the spin is of the same type.
\item Repeat step (2) for all the newly included sites, adding their neighbours with the same probability if they are also of the same type.
\item Repeat (3) until no new sites may be added.
\item Change all spins in the cluster to a randomly chosen new spin type.
\end{enumerate}
There is always a sequence of single spin flips from any
configuration to any other. The probability, $p$, is related to the reduced coupling by
(cf equation~\ref{FK})
\begin{equation}
p=\frac{e^{\beta J}-1}{e^{\beta J}}\,.
\end{equation}
It is the conditional probability that two neighbouring spins of
the same type belong to the same FK cluster. The algorithm is
therefore equivalent to isolating a single FK cluster and changing
all included spins to a new type. The benefit of using this
algorithm is that the phenomenon of critical slowing down is
eliminated. However, the average computational time required for
an update increases as the critical temperature is approached due
to increasing average cluster sizes.

This procedure describes a single update. To generate a sample, we
start with a random initial configurations of spins
which are consistent with the boundary conditions. Then the
update algorithm is run a sufficient number of times for the
energy of the system to equilibriate. In this way, we arrive at
the first sample. To calculate the number of updates required to
produce a second, independent sample, we examine the autocorrelation function:
$$
G(t_j)=\frac{1}{t_{i}}\sum_{t_{i}}{(s_1(t_{i})-\la s_1\ra
)(s_1(t_{i}+t_{j})-\la s_1\ra)}\,,\ \textrm{with}\
s_1=\sum_{i}{\delta_{s_{i},1}}\,,
$$
for a number of samples such that the average over $t_i$ yields
convergent results for large enough $t_j$. Note that $\la s_1\ra$
is also a time average. This autocorrelation function decays
exponentially as $G(t_j)\sim e^{-t_j/\tau}$. The correlation time,
$\tau$,  is a function of system size. Samples are deemed
independent after 3 correlation times.

In this paper, we consider two boundary types. The first is fixed
to spin type 1 (say). If a cluster generated by the Wolff
algorithm connects with such a boundary, we do not flip the spins.
That is to say we ignore step (5), thus keeping the boundary spins
fixed to the only allowed type. The other choice of boundary condition considered is
for the spins not to be allowed to be spin type~1 (say).
If a cluster generated by the algorithm connects with a
boundary of this type, then the allowed spin flips in step (5) of
the algorithm are constrained to  not be spin type 1. In
this way, the boundary spins fluctuate with the sample, but never
take the value~1.
\newpage

\end{document}